\documentclass[aps,prd,twocolumn,showpacs]{revtex4-1}
\usepackage{epsfig}
\usepackage{amsmath,amssymb,amsfonts,amsthm,bbm}
\usepackage[caption=false]{subfig}
\usepackage{enumerate}

\newcommand{\vek}[2]{ \left( \begin{smallmatrix} #1\\#2 \end{smallmatrix} \right) }

\newcommand{\comment}[1]{}

\renewcommand{\Im}{\text{Im}}
\renewcommand{\Re}{\text{Re}}

\begin{document}


\title{Geodesic equation in Schwarzschild--(anti-)de Sitter space--times: \\ Analytical solutions and applications}

\author{Eva Hackmann}
\email{hackmann@zarm.uni-bremen.de}
\affiliation{ZARM, University of Bremen, Am Fallturm, 28359 Bremen, Germany}

\author{Claus L{\"a}mmerzahl}
\email{laemmerzahl@zarm.uni-bremen.de}
\affiliation{ZARM, University of Bremen, Am Fallturm, 28359 Bremen, Germany}

\date{April 17, 2008}

\begin{abstract}
The complete set of analytic solutions of the geodesic equation in a Schwarzschild--(anti-)de Sitter space--time is presented. The solutions are derived from the Jacobi inversion problem restricted to the set of zeros of the theta function, called the theta divisor. In its final form the solutions can be expressed in terms of derivatives of Kleinian sigma functions. The different types of the resulting orbits are characterized in terms of the conserved energy and angular momentum as well as the cosmological constant. Using the analytical solution, the question whether the cosmological constant could be a cause of the Pioneer Anomaly is addressed. The periastron shift and its post--Schwarzschild limit is derived. The developed method can also be applied to the geodesic equation in higher dimensional Schwarzschild space--times. 
\end{abstract}



\maketitle


\section{Introduction and motivation}

All solar system observations and almost all other observations related to gravity are perfectly described within Einstein's General Relativity. This includes light deflection, the perihelion shift of planets, the gravitational time--delay (Shapiro effect) the Lense--Thirring and the Schiff effect related to the gravitomagnetic field, as well as strong field effects governing the dynamics of binary systems and, in particular, binary pulsars \cite{Will93,Will01,Krameretal06a}. However, there are two phenomena which do not fit into this scheme and still represent a mystery; that is dark matter and dark energy. Dark matter has been introduced to explain the galactic rotation curves, gravitational lensing, or particular structures in the cosmic microwave background. Dark energy is needed to describe the accelerated expansion of the universe. All related observations like the fluctuations in the cosmic microwave background, structure formation, and SN Ia are consistently described by an additional energy--momentum component which appears in the Einstein field equation as a an additional cosmological term
\begin{equation}
R_{\mu\nu} - \frac{1}{2} R g_{\mu\nu} + \Lambda g_{\mu\nu} = \kappa T_{\mu\nu} \label{EinsteinEquation}
\end{equation}
where $\Lambda$ is the cosmological constant which, using the independent observations mentioned above, has a value of $\vert \Lambda \vert \leq 10^{-52}\;{\rm m}^{-2}$. 

As a consequence, it is necessary in principle to describe all observations related to gravity within a framework including the cosmological constant. However, due to the smallness of the cosmological constant it seems unlikely that this quantity will have a large effect on smaller, that is, on solar system scales. In fact, it has been shown within an approximation scheme based on the frame given by the Schwarzschild--de Sitter space--time that the cosmological constant plays no role in all the solar system observations and also not in strong field effects \cite{KagramanovaKunzLaemmerzahl06,JetzerSereno06}. Also within a rotating version of this solution, the Kerr--de Sitter solution, no observable effects arise \cite{KerrHauckMashhoon03}. Nevertheless, there has been some discussion on whether the Pioneer anomaly, the unexplained acceleration of the Pioneer 10 and 11 spacecraft toward the inner solar system of $a_{\rm Pioneer} = (8.47 \pm 1.33) \times 10^{-10}\;{\rm m/s}^2$ \cite{Andersonetal02} which is of the order of $c H$ where $H$ is the Hubble constant, may be related to the cosmological expansion and, thus, to the cosmological constant. The same order of acceleration is present also in the galactic rotation curves which astonishingly successfully can be modeled using a modified Newtonian dynamics involving an acceleration parameter $a_{\rm MOND}$ which again is of the order of $10^{-9}\;{\rm m/s^2}$. Because of this mysterious coincidence of characteristic accelerations appearing at different scales and due to the fact that all these phenomena appear in a weak gravity or weak acceleration regime, it might be not clear whether current approximation schemes hold. This is one motivation to try to solve the equations of motion of test particles in space--times with cosmological constant analytically. 

Furthermore, by looking at the effective potential of a point particle moving in the Schwarzschild--de Sitter space--time it can be seen that for a certain range of orbital parameters a ``switching on'' of the cosmological constant may result in a dramatic change of the orbital shape: for a positive cosmological constant bound orbits may become escape orbits and for a negative cosmological constant escape orbits will become bound orbits, see Fig.~\ref{Fig:SchwDeSitterEffPot}. The characteristic distance where this happens is given by $\Lambda^{- 1/2}$ which is of the order of 5 Gpc which is roughly the radius of the visible universe and, thus, far outside the solar system and our galaxy \cite{BalagueraBoehmerNoeakowski06}. However, an orbit which is near to the separatrix of Schwarzschild geodesics may have a larger sensitivity to a cosmological constant which perhaps may not be accounted for to the required accuracy in a perturbative approach. In other words, it might be that a comparatively large acceleration $c H \sim a_{\rm Pioneer} \sim a_{\rm MOND}$ at solar system or galactic distances may be the result of a very small cosmological constant. Therefore, a definite answer to this question can be given with the help of an analytical solution only. In addition, the orbits of the Pioneer spacecrafts had been reconstructed using orbit determination programs relying on the first order post--Newtonian approximation. The difference between this approximation and the exact orbits with cosmological constant may be even more pronounced. 

There is further interest to understand explicitly the structure of geodesics in the background of black holes in anti-de Sitter space in the context of string theory and the AdS/CFT correspondence. In addition, recently there also has been a lot of work dealing with geodesics and integrability in black hole backgrounds in higher dimensions in the presence of a cosmological constant \cite{KunduriLucietti05,VasudevanStevensPage05,Vasudevan05,Chongetal05,Pageetal07}. 

Besides these physically motivated reasons, it is also of mathematical importance to derive an explicit analytical solution of the geodesic equation in a Schwarzschild--de Sitter space--time. Orbits of particles and light rays have long been used to discuss the properties of solutions of Einsteins field equations. In fact, the observation of light and particles is the only way to explore the gravitational field. All solutions of the geodesic equation in a Schwarzschild gravitational field have been presented in a seminal paper of Hagihara \cite{Hagihara31}. The solution is given in terms of the Weierstrass $\wp$--function. With the same mathematical tools one can solve the geodesic equation in a Reissner--Nordstr\"om space-time \cite{Chandrasekhar83}. The analytic solutions of the geodesic equation in a Kerr and Kerr--Newman space--time have also been given analytically (see \cite{Chandrasekhar83} for a survey). Here we expose for the first time the complete elaboration of the analytic solution of a point particle moving in a Schwarzschild--(anti) de Sitter space--time presented in \cite{HackmannLaemmerzahl08}. Also the entire set of possible solutions is described and characterized. For a specialized case orbits in a Schwarzschild--(anti) de Sitter space--time have been presented \cite{CruzOlivaresVillanueva05}.

Here we consider the general case of geodesics in the gravitational field of a spherically symmetric mass in a universe with cosmological constant $\Lambda$ (of any value), described by the Schwarzschild--(anti) de Sitter space--time. Because of the static metric and the spherical symmetry of the problem, the geodesic equation reduces to one ordinary differential equation which can be integrated formally by means of a hyperelliptic integral. Here we explicitly solve this integral. Our calculations are based on the mathematically very interesting inversion problem of hyperelliptic Abelian integrals studied first by Jacobi, Abel, Riemann, Weierstrass, and Baker in the 19th century \cite{Abel1828, Jacobi1832, Weierstrass1854, Baker1895}. The general ansatz was stated by Kraniotis and Whitehouse \cite{KraniotisWhitehouse03} and Drociuk \cite{Drociuk02} (see also \cite{Kraniotis04}), but in addition to these considerations we explicitly solve the equations of motion by restricting the problem to the set of zeros of the theta function, the so--called theta divisor. This procedure makes it possible to obtain a one--parameter solution of the, in our case, two--parameter inversion problem. This procedure was suggested by Enolskii, Pronine, and Richter \cite{EnolskiiPronineRichter03} who applied this method to the problem of the double pendulum. The resulting orbits are classified in terms of the energy and the angular momentum of the test particle as well as of the value of the cosmological constant. A detailed discussion of the resulting orbits is given. 

The found analytical solution then is applied to the question whether the cosmological constant might be the origin of the anomalous acceleration of the Pioneer spacecraft. Over the whole mission, the influence of the cosmological constant leads to a modification in the orbit of the Pioneers of the order of $10^{-4} \rm m$ only. The found solution is also used to derive the exact post--Schwarzschild approximation of the periastron shift. We also give one example for the application of this method to analytically solve the geodesic equation in higher dimensional Schwarzschild, Schwarzschild--(anti-)de Sitter or Reissner--Nordstr\"om--(anti-)de Sitter space--times. 

\section{The geodesic equation}

\begin{figure}[t]
\begin{center}
\includegraphics[width=0.45\textwidth]{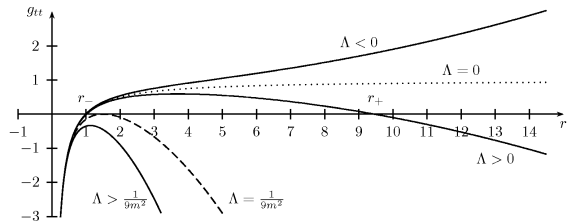}
\end{center}
\caption{The $tt$--component of the Schwarzschild--de Sitter metric for various values for $\Lambda$. The dotted line corresponds to the Schwarzschild metric. For $0 < \Lambda < 1/(9 m^2)$ there are two horizons. The dashed line corresponds to the extremal Schwarzschild--de Sitter space--time where the  two horizons coincide. For $r < r_-$ and $r > r_+$ the radial coordinate becomes timelike. \label{Fig:SdSgtt}}
\end{figure}

We consider the geodesic equation
\begin{equation}
0 = \frac{d^2 x^\mu}{ds^2} + \left\{\begin{smallmatrix} \mu \\ \rho\sigma \end{smallmatrix}\right\} \frac{dx^\rho}{ds} \frac{dx^\sigma}{ds} 
\end{equation}
where $ds^2 = g_{\mu\nu} dx^\mu dx^\nu$ is the proper time along the geodesics and 
\begin{equation}
\left\{\begin{smallmatrix} \mu \\ \rho\sigma \end{smallmatrix}\right\} = \frac{1}{2} g^{\mu\nu} \left(\partial_\rho g_{\sigma\nu} + \partial_\sigma g_{\rho\nu} - \partial_\nu g_{\rho\sigma}\right)
\end{equation}
is the Christoffel symbol, in a space--time given by the metric
\begin{align}\label{metric}
ds^2 & = \left(1 - \frac{r_{\rm S}}{r} - \frac{1}{3} \Lambda r^2 \right) dt^2 - \left(1 - \frac{r_{\rm S}}{r} - \frac{1}{3} \Lambda r^2 \right)^{-1} dr^2\nonumber\\
& \quad - r^2 (d\theta^2 + \sin^2\theta d\varphi) \, ,
\end{align}
which describes the spherically symmetric vacuum solution of (\ref{EinsteinEquation}). This Schwarzschild--de Sitter metric is characterized by the Schwarzschild--radius $r_{\rm S} = 2 M$ related to the mass $M$ of the gravitating body, and the cosmological constant $\Lambda$ (unless stated otherwise we use units where $c = G = 1$). The main features of this metric depending on the value of the cosmological constant $\Lambda$ are shown in Fig.~\ref{Fig:SdSgtt}. For a general discussion of this metric, see e.g. \cite{Rindler01,Geyer80}. The geodesic equation has to be supplemented by the normalization condition $g_{\mu\nu} \frac{dx^\mu}{ds} \frac{dx^\nu}{ds} = \epsilon$ where for massive particles $\epsilon = 1$ and for light $\epsilon = 0$.   

Because of the spherical symmetry we can restrict our consideration to the equatorial plane. Furthermore, due to the conserved energy and angular momentum 
\begin{eqnarray}
E & = & g_{tt} \frac{dt}{ds} = \left(1 - \frac{r_{\rm S}}{r} - \frac{1}{3} \Lambda r^2\right) \frac{dt}{ds} \,,\\
L & = & r^2 \frac{d\varphi}{ds}\,,
\end{eqnarray}
the geodesic equation reduces to one ordinary differential equation
\begin{equation}\label{drdvarphi}
\left(\frac{dr}{d\varphi}\right)^2 = \frac{r^4}{L^2} \left(E^2 - \left(1 - \frac{r_{\rm S}}{r} - \frac{1}{3} \Lambda r^2 \right) \left(\epsilon+\frac{L^2}{r^2} \right)\right) \,.
\end{equation}
Together with energy and angular momentum conservation we obtain the corresponding equations for $r$ as functions of $s$ and $t$
\begin{eqnarray}
\left(\frac{dr}{ds}\right)^2 & = & E^2 - \left(1 - \frac{r_{\rm S}}{r} - \frac{1}{3} \Lambda r^2 \right) \left(\epsilon+\frac{L^2}{r^2} \right)   \label{drds}\,, \\
\left(\frac{dr}{dt}\right)^2 & = & \frac{1}{E^2} \left(1 - \frac{r_{\rm S}}{r} - \frac{1}{3} \Lambda r^2\right)^2 \nonumber\\
& & \times \left(E^2 - \left(1 - \frac{r_{\rm S}}{r} - \frac{1}{3} \Lambda r^2 \right) \left(\epsilon+\frac{L^2}{r^2} \right)\right) \, .   \label{drdt}
\end{eqnarray}
Equations (\ref{drdvarphi})-(\ref{drdt}) give a complete description of the dynamics. 

\begin{figure}[t]
\begin{center}
\includegraphics[width=0.45\textwidth]{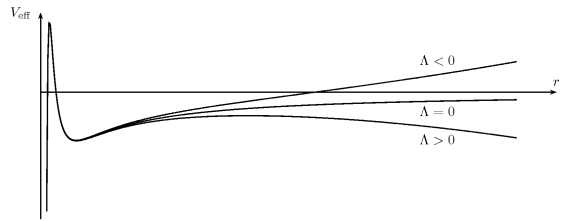}
\end{center}
\caption{The effective potential of a point particle with some given $L$ in a Schwarzschild--de Sitter space--time for different cosmological constants. \label{Fig:SchwDeSitterEffPot}}
\end{figure}

Equation (\ref{drds}) suggests the introduction of an effective potential
\begin{equation}\label{potential}
V_{\rm eff} = \frac{1}{2} \left(- \frac{1}{3} \Lambda L^2 - \epsilon \frac{r_{\rm S}}{r} + \frac{L^2}{r^2} - \frac{r_{\rm S} L^2}{r^3} - \frac{\epsilon}{3} \Lambda r^2\right)
\end{equation}
shown in Fig. \ref{Fig:SchwDeSitterEffPot}. It is worthwhile to note that for light, i.e. $\epsilon = 0$, the cosmological constant just gives a constant contribution to the effective potential and, thus, does not influence \eqref{drds} and \eqref{drdvarphi}. However, it still influences the motion of light through the timing formula \eqref{drdt}. 

As usual, we introduce a new variable $u = r_{\rm S}/r$ and obtain
\begin{equation}
\left( \frac{du}{d\varphi} \right)^2 = u^3 - u^2 + \epsilon \lambda u + \left( \lambda(\mu-\epsilon) + \rho \right) + \epsilon  \lambda \rho \frac{1}{u^2}  \label{Dgl_order3}
\end{equation}
with the dimensionless parameters
\begin{equation}\label{parameter}
\lambda := \frac{r_{\rm S}^2}{L^2} \,, \quad \mu := E^2 \quad \text{and} \quad \rho:= \frac{1}{3} \Lambda r_{\rm S}^2 \, .
\end{equation}
We rewrite \eqref{Dgl_order3} as
\begin{equation}\label{Dgl_order5}
\left( u \frac{du}{d\varphi} \right)^2 = P_5(u)
\end{equation}
with
\begin{equation}
P_5(u) := u^5 - u^4 + \epsilon \lambda u^3 + \left( \lambda(\mu-\epsilon) + \rho \right) u^2 + \epsilon \lambda \rho \, . \label{P5}
\end{equation}
If not stated otherwise, we take $\epsilon = 1$ in the following. Note that $\mu \geq 0$ and $\lambda \geq 0$. 

A separation of variables in $\eqref{Dgl_order5}$ yields
\begin{equation}\label{integralgl}
\varphi - \varphi_0 = \int_{u_0}^u \frac{u' du'}{\sqrt{P_5(u')}}\,,
\end{equation}
where $u_0 = u(\varphi_0)$. In solving integral \eqref{integralgl} there are two major issues which have to be addressed. First, the integrand is not well defined in the complex plane because of the two branches of the square root. Second, the solution $u(\varphi)$ should not depend on the integration path. If $\gamma$ denotes some closed integration path and 
\begin{equation}
\oint_\gamma \dfrac{u du}{\sqrt{P_5(u)}} = \omega \label{DefPeriod}
\end{equation}
this means that
\begin{equation}
\varphi - \varphi_0 - \omega = \int_{u_0}^u \frac{u' du'}{\sqrt{P_5(u')}}
\end{equation}
should be valid, too. Hence, the solution $u(\varphi)$ of our problem has to fulfill
\begin{equation}\label{period}
u(\varphi) = u(\varphi-\omega)
\end{equation}
for every $\omega \neq 0$ obtained from an integration \eqref{DefPeriod}. A function $u$ with the property \eqref{period} is called a periodic function with period  $\omega$. These two issues can be solved if we consider Eq.~\eqref{integralgl} to be defined on the Riemann surface $X$ of the algebraic function $x \mapsto \sqrt{P_5(x)}$. 

\section{The inversion problem}

\begin{figure}[t]
\begin{center}
\includegraphics[width=0.5\textwidth]{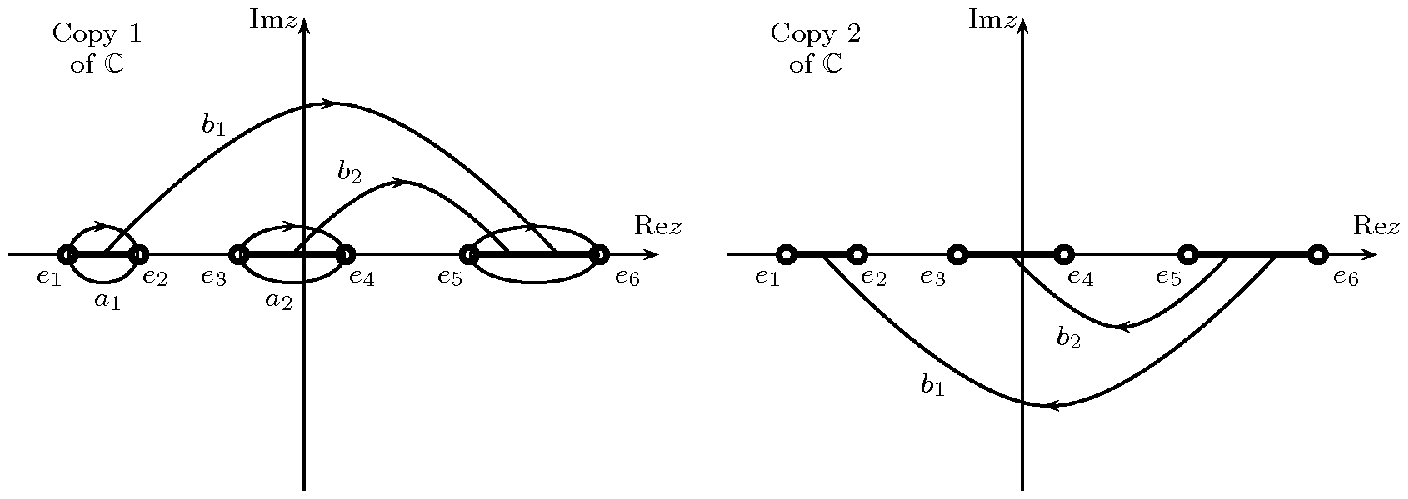}
\\
\includegraphics[width=0.4\textwidth]{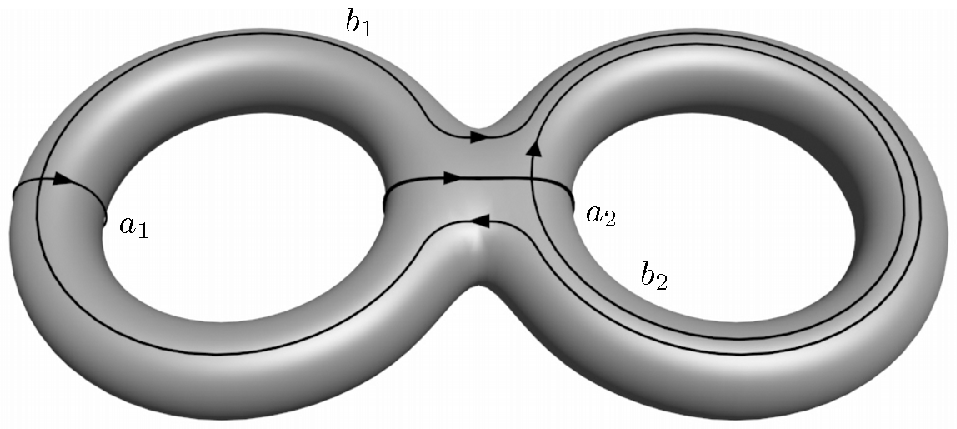}
\end{center}
\caption{Riemannian surface of genus $g=2$, with real branch points $e_1, \ldots, e_6$. Upper figure: Two copies of the complex plane with closed paths giving a homology basis $\{ a_i,b_i \,|\, i=1,\ldots,g\}$. The branch cuts (thick solid lines) are chosen from $e_{2i-1}$ to $e_{2i}$, $i=1,\ldots,g+1$. Lower figure: The "pretzel" with the topologically equivalent homology basis.}
\label{fig:brezel}
\end{figure}

Let $X$ be the compact Riemannian surface of the algebraic function $x \mapsto \sqrt{P_5(x)}$. It can be represented as the algebraic curve
\begin{equation}
X := \{ z=(x,y) \in \mathbb{C}^2 \,|\, y^2 = P_5(x) \}
\end{equation}
\cite{Miranda95} or as the analytic continuation of $\sqrt{P_5}$. The last one can be realized as a two-sheeted covering of the Riemann sphere which can be constructed in the following way: let $e_i$, $i=1,\ldots,5$, be the zeros of $P_5$ and $e_6=\infty$ (for a polynomial of $6^{\rm th}$ order the zero $e_6$ is finite). These are the so-called branch points. Now take two copies of the Riemann sphere, one for each of the two possible values of $\sqrt{P_5}$, and cut them between every two of the branch points $e_i$ in such a way that the cuts do not touch each other. These are the so-called branch cuts, see Fig.~\ref{fig:brezel}. Of course, the two copies have to be identified at the branch points where the two values of $\sqrt{P_5}$ are identical. They are then glued together along the branch cuts in such a way that $\sqrt{P_5}$ together with all its analytic continuations is uniquely defined on the whole surface. On this surface $x \mapsto \sqrt{P_5(x)}$ is now a single--valued function. This construction can be visualized as a "pretzel", see Fig.~\ref{fig:brezel}. For a strict mathematical description of the construction of a compact Riemannian surface, see \cite{RauchFarkas74}, for example. 

Every compact Riemannian surface can be equipped with a homology basis $\{ a_i,b_i \,|\, i=1,\ldots,g\} \in H_1(X,\mathbb{Z})$ of closed paths as shown in Fig.~\ref{fig:brezel}, where $g$ is the genus of the Riemannian surface, see the next section. From the construction of the Riemannian surface it is already clear that integrals over these closed paths indeed do not evaluate to zero and, hence, have to be periods of the solution of \eqref{integralgl}. The task now is to analyze the details of periodic functions on such Riemannian surfaces.

\subsection{Preliminaries}

Compact Riemannian surfaces are characterized by their genus $g$. This can be defined as the dimension of the space of holomorphic differentials on the Riemannian surface or, topologically seen, as the number of `holes' of the Riemannian surface. Let $P_d = \sum_{s = 0}^d \lambda_s x^s$ be a polynomial of degree $d$ with only simple zeros and $X$ be the Riemann surface of $\sqrt{P_d}$. Then the genus of $X$ is equal to $g = \left[\frac{d-1}{2}\right]$, where $[x]$ denotes the largest integer less or equal than $x$ \cite{BuchstaberEnolskiiLeykin97}. Hence, in our case of $P_5$ the genus of the Riemann surface is $g=2$.

In order to construct periodic functions on a Riemann surface we first have to define a canonical basis of the space of holomorphic differentials $\{ dz_i \,|\, i=1,\ldots,g\}$ and of associated meromorphic differentials $\{dr_i \,|\, i=1,\ldots,g\}$ on the Riemann surface by
\begin{align}
dz_i & := \frac{x^{i-1} dx}{\sqrt{P_d(x)}} \label{holomorphdiff} \,,\\ 
dr_i & := \sum_{k=i}^{2g+1-i} (k+1-i) \lambda_{k+1+i} \frac{x^k dx}{4 \sqrt{P_d(x)}} \, , \label{meromorphicdifferentials}
\end{align}
with $\lambda_j$ being the coefficients of the polynomial $P_d$ \cite{BuchstaberEnolskiiLeykin97}. In our case these differentials are given by
\begin{align}
dz_1 & := \frac{dx}{\sqrt{P_5(x)}}\,, & dz_2 &:= \frac{x dx}{\sqrt{P_5(x)}}\,, \\
dr_1 &:= \frac{3x^3-2x^2+\lambda x}{4 \sqrt{P_5(x)}} dx \,, & dr_2 &:= \frac{x^2 dx}{4 \sqrt{P_5(x)}}\,,
\end{align}
where $\lambda$ is defined in \eqref{parameter}. We also introduce the period matrices $( 2\omega, 2\omega^\prime )$ and  $( 2\eta, 2\eta^\prime )$ related to the homology basis
\begin{equation}\label{periodmatrices}
\begin{aligned}
2 \omega_{ij} &:= \oint_{a_j} dz_i\,, & \qquad 2 \omega'_{ij} &:= \oint_{b_j} dz_i\,, \\
2 \eta_{ij} &:= - \oint_{a_j} dr_i\,, & \qquad 2 \eta'_{ij} &:= - \oint_{b_j} dr_i\, .
\end{aligned}
\end{equation}
The differentials in \eqref{holomorphdiff} and \eqref{meromorphicdifferentials} have been chosen such that the components of their period matrices fulfill the Legendre relation
\begin{equation}
\begin{pmatrix} \omega & \omega'\\ \eta & \eta' \end{pmatrix} \begin{pmatrix} 0 & - \mathbbm{1}_g \\ \mathbbm{1}_g & 0 \end{pmatrix}   \begin{pmatrix} \omega & \omega'\\ \eta & \eta' \end{pmatrix}^t = - \frac{1}{2} \pi i \begin{pmatrix} 0 & - \mathbbm{1}_g \\\mathbbm{1}_g & 0 \end{pmatrix} \,,
\end{equation}
where $\mathbbm{1}_g$ is the $g \times g$ unit matrix, \cite{BuchstaberEnolskiiLeykin97}. 

Finally we introduce the normalized holomorphic differentials 
\begin{equation}
d\vec{v} := (2 \omega)^{-1} d\vec{z} \, , \qquad d\vec z = \begin{pmatrix} dz_1 \\ dz_2 \\ \vdots \\ dz_g \end{pmatrix} \, .
\end{equation}
The period matrix of these differentials is given by $(\mathbbm{1}_g,\tau)$, where $\tau$ is defined by
\begin{equation}
\tau:=  \omega^{-1} \omega'\,. \label{normalizedtau}
\end{equation}
It can be shown \cite{Mumford83} that this normalized matrix $\tau$ always is a Riemannian matrix, that is, $\tau$ is symmetric and its imaginary part $\Im \tau$ is positive definite.

\subsection{Jacobi's inversion problem}

Let us consider now the Abel map
\begin{equation}
\mathcal{A}_{x_0}: X \to {\rm Jac}(X)\,, \quad x \mapsto \int_{x_0}^x d\vec{z}
\end{equation}
from the Riemannian surface $X$ to the Jacobian ${\rm Jac}(X) = \mathbb{C}^g/\Gamma$ of $X$, where $\Gamma=\{ \omega v + \omega^\prime v^\prime \mid v, v^\prime \in \mathbb{Z}^g \}$ is the lattice of periods of the differential $d\vec{z}$. The image $\mathcal{A}_{x_0}(X)$ of $X$ by this continuous function is of complex dimension one and, thus, an inverse map $\mathcal{A}_{x_0}^{-1}$ is not defined for all points of ${\rm Jac}(X)$. However, the $g$--dimensional Abel map
\begin{equation}\label{Abel}
A_{x_0} : S^gX \to {\rm Jac}(X)\,, \quad (x_1,\ldots,x_g)^t \mapsto \sum_{i=1}^g \int_{x_0}^{x_i} d\vec{z}
\end{equation}
from the $g$th symmetric product $S^gX$ of $X$ (the set of unordered "vectors" $(x_1,\ldots, x_g)^t$ where $x_i \in X$) to the Jacobian is one--to--one almost everywhere. Jacobi's inversion problem is now to determine $\vec x$ for given $\vec \varphi$ from the equation 
\begin{equation}\label{Jacobigen}
\vec{\varphi} = A_{x_0} ( \vec{x} )\,.
\end{equation}
In our case $g=2$ this reads
\begin{equation}\label{Jacobi}
\begin{split}
\varphi_1  & = \int_{x_0}^{x_1} \frac{dz}{\sqrt{P_5(z)}} + \int_{x_0}^{x_2} \frac{dz}{\sqrt{P_5(z)}} \,, \\ 
\varphi_2  & = \int_{x_0}^{x_1} \frac{z dz}{\sqrt{P_5(z)}} + \int_{x_0}^{x_2} \frac{z dz}{\sqrt{P_5(z)}} \,.
\end{split}
\end{equation}
We will see later, that we can solve our problem \eqref{integralgl} as a limiting case of this Jacobi inversion problem.

\subsection{Theta Functions}

The Riemannian surface of genus $g$ has $2g$ independent closed paths, each corresponding to a period of the functions defined on these surfaces and, hence, to a period of the solution $u$ of \eqref{integralgl}. In order to construct $2g$--periodic functions, we need the theta function $\vartheta : \mathbb{C}^g \to \mathbb{C}$, 
\begin{equation}
\vartheta(\vec z;\tau) := \sum_{\vec m \in {\mathbb{Z}}^g} e^{i \pi \vec m^t (\tau \vec m + 2 \vec z)}\,.
\end{equation}
The series on the right-hand side converges absolutely and uniformly on compact sets in $\mathbbm{C}^g$ and, thus, defines a holomorphic function in $\mathbbm{C}^g$. This is obvious from the estimate $\Re( \vec m^t (i\tau) \vec m^t ) \leq -c \vec m^t \vec m$ for some constant $c > 0$, what follows from the fact that $\Re( i \tau)$ is negative definite. The theta function is already periodic with respect to the columns of $\mathbbm{1}_g$ and quasiperiodic with respect to the columns of $\tau$, i.e., for any $n \in \mathbb{Z}^g$ the relations
\begin{align}
\vartheta(\vec z + \mathbbm{1}_g \vec n;\tau) & = \vartheta(\vec z;\tau)\,,\\
\vartheta(\vec z + \tau \vec n;\tau) & = e^{- i \pi \vec n^t (\tau \vec n + 2 \vec z)} \vartheta(\vec z;\tau)
\end{align} 
hold. We will also need the theta function with characteristics\footnote{The symbol $\frac{1}{2} \mathbbm{Z}^g$ denotes the set of all $g$-dimensional vectors with half--integer entries $\ldots, -\frac{3}{2}, -1, -\frac{1}{2}, 0, \frac{1}{2}, 1, \frac{3}{2}, \ldots$} $\vec g, \vec h \in \frac{1}{2} \mathbbm{Z}^g$ defined by
\begin{align}
\vartheta[\vec g, \vec h](\vec z;\tau) &:= \sum_{\vec m \in \mathbb{Z}^g} e^{i \pi (\vec m + \vec g)^t (\tau (\vec m + \vec g) + 2 \vec z + 2 \vec h)} \nonumber\\
& = e^{i \pi \vec g^t (\tau \vec g + 2 \vec z + 2 \vec h)} \vartheta(\vec z + \tau \vec g + \vec h; \tau)\,. \label{thetach}
\end{align}
Later it will be important that for every $\vec g, \vec h$ the set $\Theta_{\tau \vec g+ \vec h} := \{\vec z \in {\mathbb{C}}^g \mid \vartheta[\vec g, \vec h](\vec z;\tau) = 0\}$, called a {\it theta divisor}, is a $(g-1)$--dimensional subset of ${\rm Jac}(X)$, see \cite{Mumford83} or \eqref{thetadiv}.

The solution of Jacobi's inversion problem \eqref{Jacobi} can be explicitly formulated in terms of functions closely related to the theta function. First, consider the Riemann theta function 
\begin{equation}
\vartheta_e (x;\tau) := \vartheta \left( \int_{x_0}^x d\vec{v} - \vec e; \tau \right) \, ,
\end{equation}
with some arbitrary but fixed $\vec e \in {\mathbb{C}}^g$. 
The Riemann vanishing theorem, see e.g.~\cite{Mumford83}, states that the Riemann theta function is either identically to zero or has exactly $g$ zeros $x_1,\ldots,x_g$ for which 
\begin{equation}
\sum_{i=1}^g \int_{x_0}^{x_i} d\vec{v} = \vec e + \vec K_{x_0}
\end{equation}
holds (modulo periods). Here $\vec K_{x_0} \in \mathbbm{C}^g$ is the vector of Riemann constants with respect to the base point $x_0$ given by ($\tau_{jj}$ is the $j$th diagonal element) 
\begin{equation}
K_{x_0, j} = \frac{1+\tau_{jj}}{2} - \sum_{l\neq j} \oint_{a_l} \left( \int_{x_0}^x dv_j \right) dv_l(x)\,.
\end{equation}
If the base point $x_0$ is equal to $\infty$, this vector can be determined by  
\begin{equation}\label{Riemann_Konst}
\vec{K}_{\infty} = \sum_{i=1}^g \int_{\infty}^{e_{2i}} d\vec{v}\,, 
\end{equation}
where $e_{2i}$ is the starting point of one of the branch cuts not containing $\infty$ for each $i$, see \cite{BuchstaberEnolskiiLeykin97}. Hence, $\vec K_{\infty}$ can be expressed as a linear combination of half-periods in this case. For problems of hyperelliptic nature it is usually assumed that the Riemann theta function $\vartheta_e$ does not vanish identically. However, here we are interested in the opposite case: we want to restrict Jacobi's inversion problem \eqref{Jacobi} to the set of zeros of $\vartheta( \cdot + \vec K_{x_0}; \tau)$, which is called the theta divisor $\Theta_{\vec K_{x_0}}$.

The solution of \eqref{Jacobi} and, thus, of \eqref{Dgl_order5} can be formulated in terms of the derivatives of the Kleinian sigma function $\sigma : {\mathbb{C}}^g \rightarrow \mathbb{C}$,
\begin{equation}\label{Def_sigma}
\sigma(\vec z) = C e^{- \frac{1}{2} \vec z^t \eta \omega^{-1} \vec z} \vartheta \left( (2 \omega)^{-1}\vec z + \vec K_{x_0};\tau \right)\,,
\end{equation}
where the constant $C$ can be given explicitly, see \cite{BuchstaberEnolskiiLeykin97}, but does not matter here.
Jacobi's inversion problem can be solved in terms of the second logarithmic derivative of the sigma function called the generalized Weierstrass functions
\begin{equation}\label{Def_Weier}
\wp_{ij}(\vec z) = - \frac{\partial}{\partial z_i}  \frac{\partial}{\partial z_j} \log \sigma(\vec z) = \frac{\sigma_i(\vec z) \sigma_j(\vec z) - \sigma(\vec z) \sigma_{ij}(\vec z)}{\sigma^2(\vec z)} \,,
\end{equation}
where $\sigma_{i}$ denotes the derivative of the sigma function with respect to the $i-$th component.

\subsection{The solution of the Jacobi inversion problem}

The solution of Jacobi's inversion problem \eqref{Jacobigen} can be given in terms of generalized Weierstrass functions. Let $X$ be the Riemannian surface of $\sqrt{P}$ where $P$ is defined by $P(x) := \sum_{i=0}^{2g+1} \lambda_i x^i$ (this can always be achieved by a rational transformation). Then the components of the solution vector $\vec{x}=(x_1,\ldots,x_g)^t$ are given by the $g$ solutions of
\begin{equation}
\frac{\lambda_{2g+1}}{4} x^g - \sum_{i=1}^g \wp_{gi}(\vec{\varphi}) x^{i-1} = 0\,,
\end{equation} 
where $\vec\varphi$ is the left hand side of \eqref{Jacobigen}. Since $\vec x \in S^2X$ there is no way to define an order of the components of $\vec{x}$. In our case of $g=2$, we can rewrite this result with the help of the theorems by Vieta in the form
\begin{equation}\label{Jacobisol}
\begin{split}
x_1+x_2 = & \frac{4}{\lambda_{5}} \wp_{22}(\vec \varphi)\,, \\
x_1 \, x_2 = & - \frac{4}{\lambda_{5}} \wp_{12}(\vec \varphi) \, .
\end{split}
\end{equation}

\section{Solution of the equation of motion in Schwarzschild--(anti-)de Sitter space--time}

Now we apply the results of the preceding section to the problem of the equation of motion in Schwarzschild--(anti-)de Sitter space--time, Eqs.~\eqref{Dgl_order5} and \eqref{integralgl}. As already mentioned before, the solution of the equation of motion can be found as a limiting case of the solution of Jacobi's inversion problem in the case of genus $g=2$. 

\subsection{The analytic expression}\label{solution}

To start with, we rewrite Jacobi's inversion problem \eqref{Jacobi} in the form
\begin{equation}\label{Jacobi2}
\begin{split}
\phi_1 = & \int_{\infty}^{u_1} \frac{dx}{\sqrt{P_5(x)}} + \int_{\infty}^{u_2} \frac{dx}{\sqrt{P_5(x)}} \,,\\
\phi_2 = & \int_{\infty}^{u_1} \frac{x dx}{\sqrt{P_5(x)}} + \int_{\infty}^{u_2} \frac{x dx}{\sqrt{P_5(x)}} \,,
\end{split}
\end{equation}
where 
\begin{equation}\label{defPhi}
\vec \phi = \vec \varphi - 2 \int_{u_0}^{\infty} d\vec z \,.
\end{equation} 
Note that the right-hand side of \eqref{Jacobi2} is exactly $\vec A_\infty(\vec{u})$, the image of the Abel map defined in \eqref{Abel}, i.e. $\vec \phi = \vec A_\infty(\vec u)$. We use the obvious identity (compare \cite{EnolskiiPronineRichter03})
\begin{equation}
u_1 = \lim_{u_2 \to \infty} \frac{u_1 u_2}{u_1+ u_2}
\end{equation}
and insert the solution of Jacobi's inversion problem \eqref{Jacobisol}. Then 
\begin{align}
u_1 & = - \lim_{u_2 \to \infty} \frac{\wp_{12}(\vec\phi)}{\wp_{22}(\vec\phi)} 
\nonumber\\
& = \lim_{u_2 \to \infty} \frac{\sigma(\vec\phi) \sigma_{12}(\vec\phi) - \sigma_1(\vec\phi) \sigma_2(\vec\phi)}{\sigma_2^2(\vec\phi) - \sigma \sigma_{22}(\vec\phi)} \nonumber \\
& = \frac{\sigma(\vec\phi_\infty) \sigma_{12}(\vec\phi_\infty) - \sigma_1(\vec\phi_\infty) \sigma_2(\vec\phi_\infty)}{\sigma_2^2(\vec\phi_\infty) - \sigma(\vec\phi_\infty) \sigma_{22}(\vec\phi_\infty)} \, ,\label{u_1}
\end{align}
where
\begin{equation}
\vec\phi_\infty = \lim_{u_2 \to \infty} \vec \phi = \vec A_\infty(\vec u_\infty) 
\end{equation}
with $\vec u_\infty = \vek{u_1}{\infty}$. Note that the definition of the sigma function \eqref{Def_sigma} and, hence, of the generalized Weierstrass functions \eqref{Def_Weier} includes the vector of Riemann constant $\vec{K}_{x_0}$ with $x_0=\infty$ in our case, which is given by $\vec{K}_\infty = \tau \vek{1/2}{1/2} + \vek{0}{1/2}$ (see \eqref{Riemann_Konst} or \cite{Mumford83}). 

The above limiting process also transfers Jacobi's inversion problem to the theta divisor $\Theta_{\vec K_\infty}$. With  $(2 \omega)^{-1} \vec\phi_\infty = (2 \omega)^{-1} \vec A_\infty(\vec u_\infty) = \int_\infty^{u_1} d\vec v$ and the theorem 
\begin{align}\label{thetadiv}
& \vartheta \left[ \vek{1/2}{1/2},\vek{0}{1/2} \right] (\vec z;\tau) = 0 \nonumber\\
\Leftrightarrow & \qquad \exists x : \vec z = \int_{\infty}^{x} d\vec v 
\end{align}
proven by Mumford \cite{Mumford83} it follows that 
\begin{equation}
0 = \vartheta\left[\vek{1/2}{1/2},\vek{0}{1/2}\right]( (2 \omega)^{-1} \vec\phi_\infty ;\tau)\,.
\end{equation}
Via Eq.~\eqref{thetach} this is equivalent to 
\begin{equation}
0 = \vartheta \left( (2 \omega)^{-1} \vec\phi_\infty + \tau \vek{1/2}{1/2} + \vek{0}{1/2}; \tau\right) \,
\end{equation} 
and with \eqref{Def_sigma} this means
\begin{equation}
\sigma(\vec\phi_\infty) = 0 \, .
\end{equation}
We first use this result in \eqref{u_1} and obtain
\begin{equation}
u_1 = - \frac{\sigma_1(\vec\phi_\infty)}{\sigma_2(\vec\phi_\infty)} \, .
\end{equation}

Theorem \eqref{thetadiv} also tells us that $(2 \omega)^{-1} \vec\phi_\infty$ is an element of the theta divisor $\Theta_{\vec K_\infty}$, i.e. the set of zeros of $ \vartheta\left[\vek{1/2}{1/2},\vek{0}{1/2}\right]$, and that, in the case $g=2$, $\Theta_{\vec K_\infty}$ is a manifold of complex dimension one.
Note that the restriction to the theta divisor is only possible because $\infty$ is a branch point what is essential for the validity of theorem \eqref{thetadiv}.

Since $\Theta_{\vec K_\infty}$ is a one--dimensional subset of ${\mathbbm{C}}^2$, there is a one--to--one functional relation between the first and the second component of $(2\omega)^{-1} \vec \phi_{\infty}$. By the definition of $\vec \phi_{\infty}$ in \eqref{u_1} and Eq. \eqref{defPhi} we have
\begin{align}\label{phiinfty}
\vec \phi_{\infty} & = \lim_{u_2 \to \infty} \vec \phi \nonumber\\
& = \lim_{u_2 \to \infty} \vec \varphi  - 2 \int_{u_0}^{\infty} d\vec z \nonumber\\
& = \int_{u_0}^{u_1} d\vec z - \int_{u_0}^{\infty} d\vec z\,.
\end{align}
The physical coordinate $\varphi$ is given by \eqref{integralgl},
\begin{equation}
\varphi = \int_{u_0}^{u_1} \frac{z dz}{\sqrt{P_5(z)}} + \varphi_0 = \int_{u_0}^{u_1} dz_2 + \varphi_0\,.
\end{equation}
We insert this in \eqref{phiinfty} and obtain
\begin{align}
\vec \phi_{\infty} & = \begin{pmatrix} \int_{u_0}^{u_1} dz_1 - \int_{u_0}^{\infty} dz_1 \\ \varphi - \varphi_0 - \int_{u_0}^{\infty} dz_2 \end{pmatrix} \nonumber\\
& = \begin{pmatrix} \int_{u_0}^{u_1} dz_1 - \int_{u_0}^{\infty} dz_1 \\ \varphi - \varphi'_0 \end{pmatrix}\,,
\end{align}
where $\varphi'_0 = \varphi_0 + \int_{u_0}^{\infty} dz_2$ depends only on the initial values $u_0$ and $\varphi_0$. We choose for each $\varphi$ a $\varphi_1$ such that $\vec \varphi_\Theta := \begin{pmatrix} \varphi_1 \\ \varphi-\varphi'_0 \end{pmatrix}$ is equal to $\vec \phi_\infty$. Then $(2\omega)^{-1} \vec \varphi_\Theta = (2\omega)^{-1} \vec \phi_\infty$ is an element of the theta divisor $\Theta_{\vec K_\infty}$ and we finally obtain 
\begin{equation}
r(\varphi) = \frac{r_{\rm S}}{u(\varphi)} = - r_{\rm S} \frac{\sigma_2 (\vec\phi_\infty)}{\sigma_1(\vec\phi_\infty)} = - r_{\rm S} \frac{\sigma_2(\vec \varphi_\Theta)}{\sigma_1(\vec \varphi_\Theta)} \,.
\end{equation}
This is the analytic solution of the equation of motion of a point particle in a Schwarzschild--(anti-)de Sitter space--time. This solution is valid in all regions of the Schwarzschild--(anti-)de Sitter space--time and for both signs of the cosmological constant and can be computed with arbitrary accuracy. The explicit computation of the solution is described in Appendix A.

\subsection{Light trajectories}

In the case of light trajectories, the situation simplifies considerably. The equation of motion is then given by 
\begin{equation}
\left(\frac{du}{d\varphi}\right)^2 = u^3 - u^2 + \lambda \mu + \frac{1}{3} \rho =: P_3(u) \,.
\end{equation}
Light rays are uniquely given and, thus, uniquely characterized by the extremal distance to the gravitating body, that is, the smallest or largest distance (in the case of a Schwarzschild space--time, it is also possible - due to its asymptotic flatness - to take the impact parameter as characteristic of a light ray). This extremal distance $u_0$ is characterized by 
\begin{equation}
\left.\frac{du}{d\varphi}\right|_{u = u_0} = 0 \,,
\end{equation}
which gives $u_0^3 - u_0^2 + \lambda \mu + \frac{1}{3} \rho = 0$.
Then our equation of motion is
\begin{equation}
\left(\frac{du}{d\varphi}\right)^2 = u^3 - u^2 - u_0^3 + u_0^2 \,
\end{equation}
which is the same type of equation as in Schwarzschild geometry. With a substitution $u = 4x + \frac{1}{3}$ this reads
\begin{equation}\label{Dgl_ell}
\left(\frac{dx}{d\varphi}\right)^2 = 4 x^3 - g_2 x - g_3\,
\end{equation}
where 
\begin{equation}
g_2 := \frac{1}{12} \, , \qquad g_3:= \frac{1}{8} \left(\frac{1}{27} + \frac{1}{2} \left(u_0^3 - u_0^2\right)\right)
\end{equation}
are the Weierstrass invariants. This differential equation can be solved directly in terms of elliptic functions, i.e. 
\begin{equation}
r(\varphi) = \frac{r_{\rm S}}{u(\varphi)} = \frac{r_{\rm S}}{4 x(\varphi) + \frac{1}{3}} = \frac{r_{\rm S}}{4 \wp(\varphi-\varphi'_0; g_2,g_3) + \frac{1}{3}}\,,
\end{equation}
where $\wp$ is the Weierstrass function \cite{GradshteynRyzhik83,AbramowitzStegun68} and $\varphi'_0$ is given by the initial values $\varphi_0$ and $x_0$, $\varphi'_0 = \varphi_0 + \int_{x_0}^\infty \frac{dx}{\sqrt{P_3(x)}}$. The corresponding light trajectories have been exhaustively discussed in \cite{Hagihara31}. 

In a recent paper \cite{RindlerIshak07} Rindler and Ishak discussed light deflection in a Schwarzschild--de Sitter space--time. Though the equation of motion is the same as in Schwarzschild space--time, the measuring process for angles reintroduces the cosmological constant in the observables. According to their scheme, the exact angle between the radial direction and the spatial direction of the light ray is now given by
\begin{equation}
\tan\psi = \sqrt{\frac{1 - \frac{r_{\rm S}}{r(\varphi)} - \frac{1}{3} \Lambda r^2(\varphi)}{\left|\frac{r^2(\varphi)}{r_0^2} \left(1 - \frac{r_{\rm S}}{r_0}\right) - \left(1 - \frac{r_{\rm S}}{r(\varphi)}\right)\right|}}  \, ,
\end{equation}
where in the expression $dr/d\varphi$ from \eqref{drdvarphi} the $\frac{E^2}{L^2} + \frac{1}{3} \Lambda$ has been replaced by the $r_0$ related to $u_0$. This now is valid for all light rays, not only for those rays showing a small deflection as discussed in \cite{RindlerIshak07}. 

\section{The classification of the solutions}

\subsection{General classification}

Having an analytical solution at hand we can explore the set of all possible solutions in a systematic manner. The shape of an orbit depends on the energy $E$ and the angular momentum $L$ of the particle under consideration as well as the cosmological constant $\Lambda$ (the Schwarzschild radius has been absorbed through a rescaling of the radial coordinate). These quantities are all contained in the polynomial $P_5(u)$ through the parameters $\lambda$, $\mu $ and $\rho$ \eqref{parameter}. Since $r$ (and $u$) should be real and positive it is clear that the  physically acceptable regions are given by those $u$ for which $E > V_{\rm eff}$. The zeros of $P_5$ are related to the points of intersection of $E$ and $V_{\rm eff}$, and a real and positive $P_5$ is equivalent to $E > V_{\rm eff}$ as can also be seen from \eqref{Dgl_order5}. Hence, the number of positive real zeros of $P_5$ uniquely characterizes the form of the resulting orbit.

\begin{figure}
\subfloat[][No real positive zero]{
\includegraphics[width=0.18\textwidth]{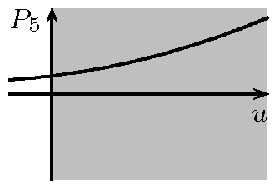}
}
\subfloat[][One real positive zero]{
\includegraphics[width=0.18\textwidth]{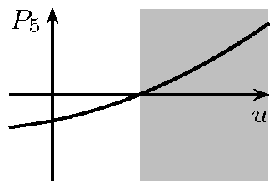}
}\\
\subfloat[][Two real positive zeros]{
\includegraphics[width=0.18\textwidth]{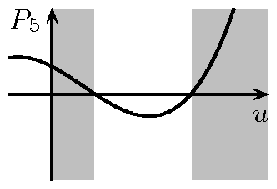}
}
\subfloat[][Three real positive zeros]{
\includegraphics[width=0.18\textwidth]{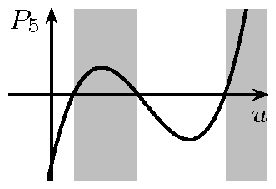}
}\\
\subfloat[][Four real positive zeros]{
\includegraphics[width=0.18\textwidth]{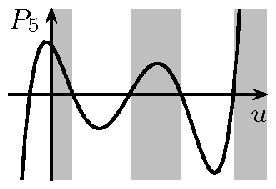}
}
\caption{The five possibilities of having real positive zeros of $P_5(u)$. The allowed regions of particle motion are shaded in gray. The zeros correspond to the zeros of $V_{\rm eff} = E$ in Fig.~\ref{Fig:SchwDeSitterEffPot} (note that $u = 0$ corresponds to $r = \infty$ and $u = \infty$ to $r = 0$). Bound nonterminating, quasiperiodic orbits exist only if there are three or more positive zeros. \label{GraphPolynomials}}
\end{figure}

Since $P_5$ goes to $-\infty$ if $x \to -\infty$ and to $\infty$ if $x \to \infty$, $P_5(0)$ is positive if the number of positive real zeros of $P_5$ is even and negative if it is odd. If we denote by $e_1,\ldots,e_n$ the positive real zeros, then it follows that the physically acceptable regions are given by $[0,e_1], [e_2,e_3], \ldots, [e_n,\infty]$ if $n$ is even and by $[e_1,e_2], \ldots, [e_n,\infty]$ if $n$ is odd. With respect to $r$ we have the following classes of orbits (see Fig. \ref{GraphPolynomials}):
\begin{enumerate}[(i)]\itemsep=-2pt
\item the region $[0,e_1]$ corresponds to escape orbits, 
\item the region $[e_n,\infty]$ corresponds to orbits falling into the singularity, i.e. to terminating orbits, and 
\item the regions $[e_i,e_{i+1}]$ correspond to bound orbits. 
\end{enumerate}
This means that for any arrangement of zeros of $P_5$ there exist terminating orbits. Furthermore, for an even number of positive real zeros we have escape orbits and for more than three real positive zeros we have bound orbits. The case that there is no positive real zero corresponds to a particle coming from infinity and falling to the singularity, see Fig.~\ref{GraphPolynomials}. Quasiperiodic bound orbits exists only if there are three or more positive zeros. 

It can be shown that there are no more than four real positive zeros for our polynomial \eqref{P5}: We decompose the polynomial $P_5$ into its (in general complex) zeros $P_5(u) = (u - u_1) (u - u_2) (u - u_3) (u - u_4) (u - u_5)$. Multiplication and comparison of the coefficients of the terms linear in $u$ yields
\begin{equation}
u_1 u_2 u_3 u_4 + u_1 u_2 u_3 u_5 + u_1 u_2 u_4 u_5 + u_1 u_3 u_4 u_5 + u_2 u_3 u_4 u_5 = 0 \,. \label{P5_zeros}
\end{equation}
The assumption that all zeros are real and positive contradicts Eq. \eqref{P5_zeros}. Therefore, in any case there are at most four real positive zeros. 

\begin{figure*}[t!]
\subfloat[][$\Lambda = -10^{-5} \text{km}^{-2}$]{
\includegraphics[width=0.31\textwidth]{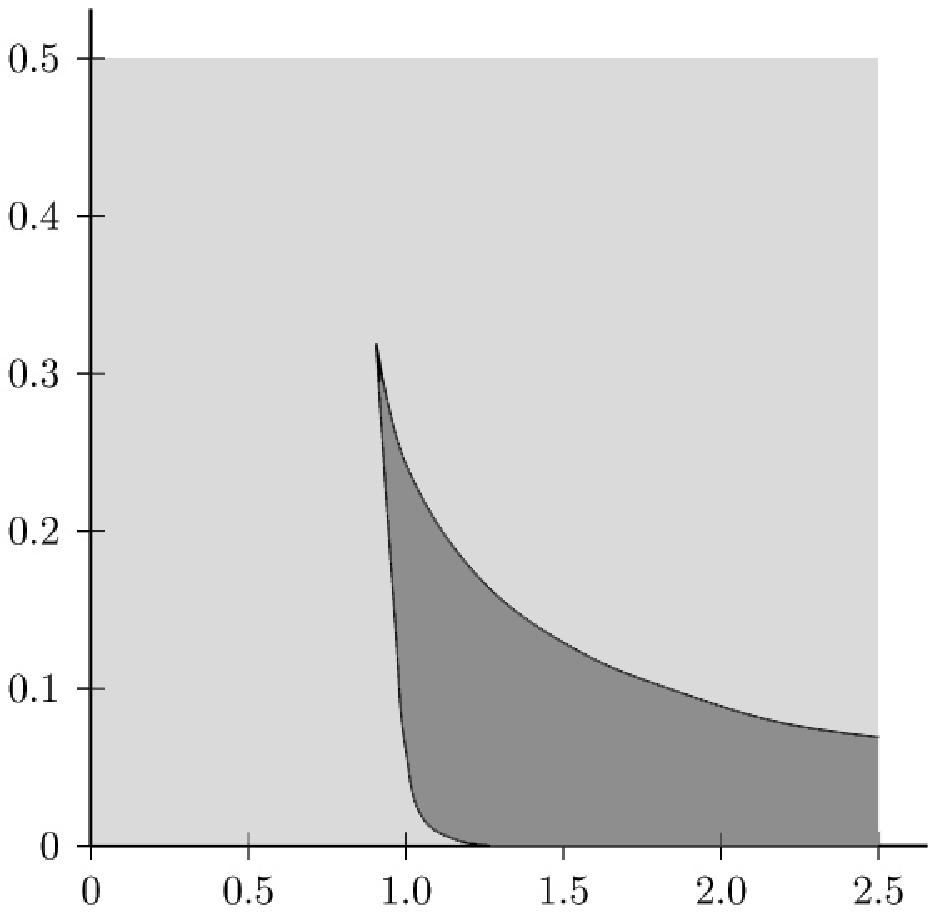}
}
\subfloat[][$\Lambda = -10^{-10} \text{km}^{-2}$]{
\includegraphics[width=0.31\textwidth]{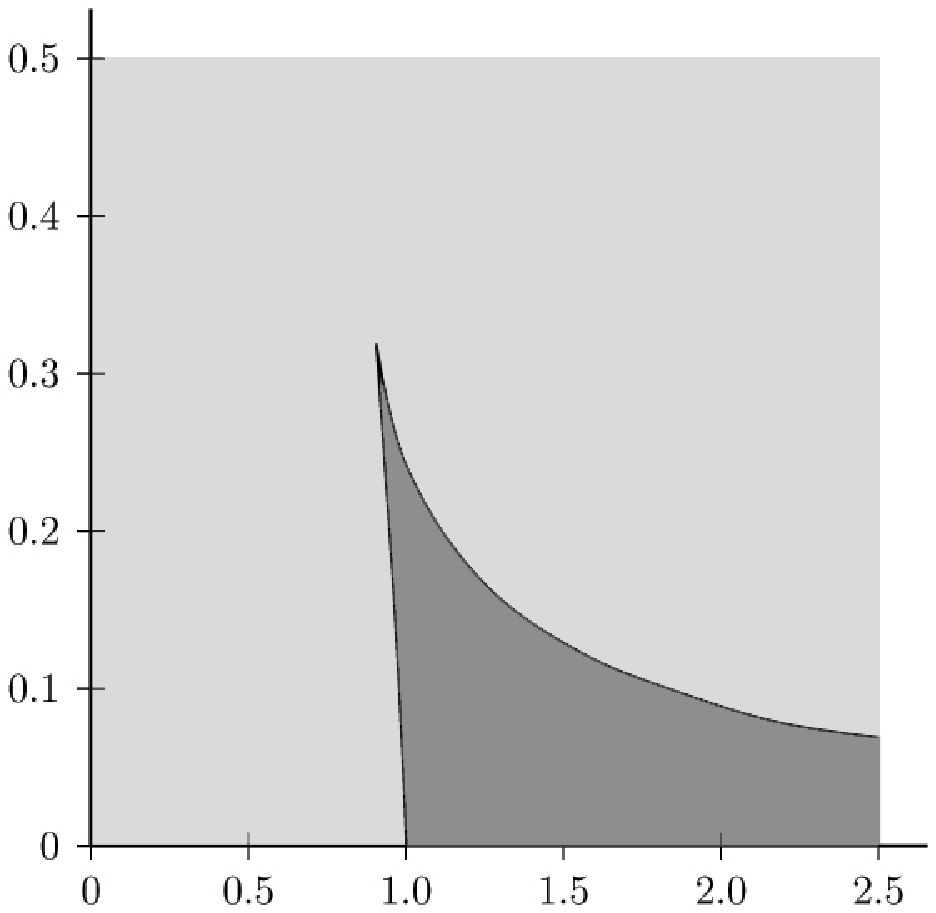}
}
\subfloat[][$\Lambda = -10^{-45} \text{km}^{-2}$]{
\includegraphics[width=0.31\textwidth]{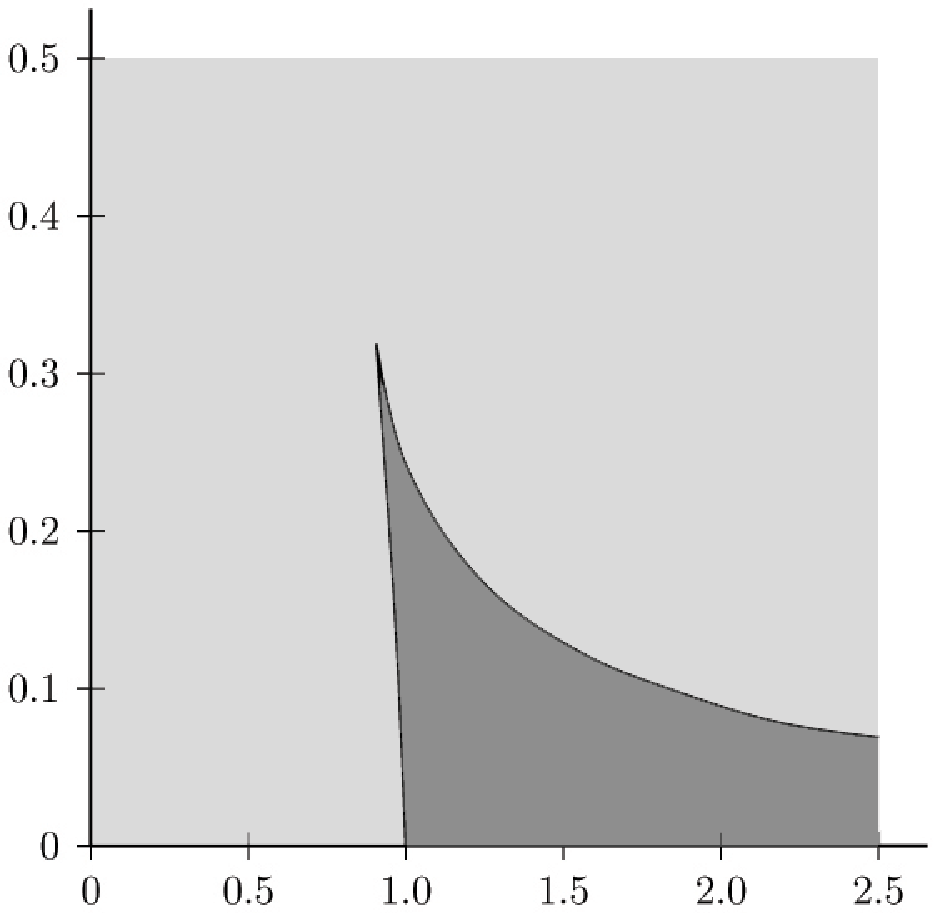}
} \\
\subfloat[][$\Lambda = 0$]{
\includegraphics[width=0.31\textwidth]{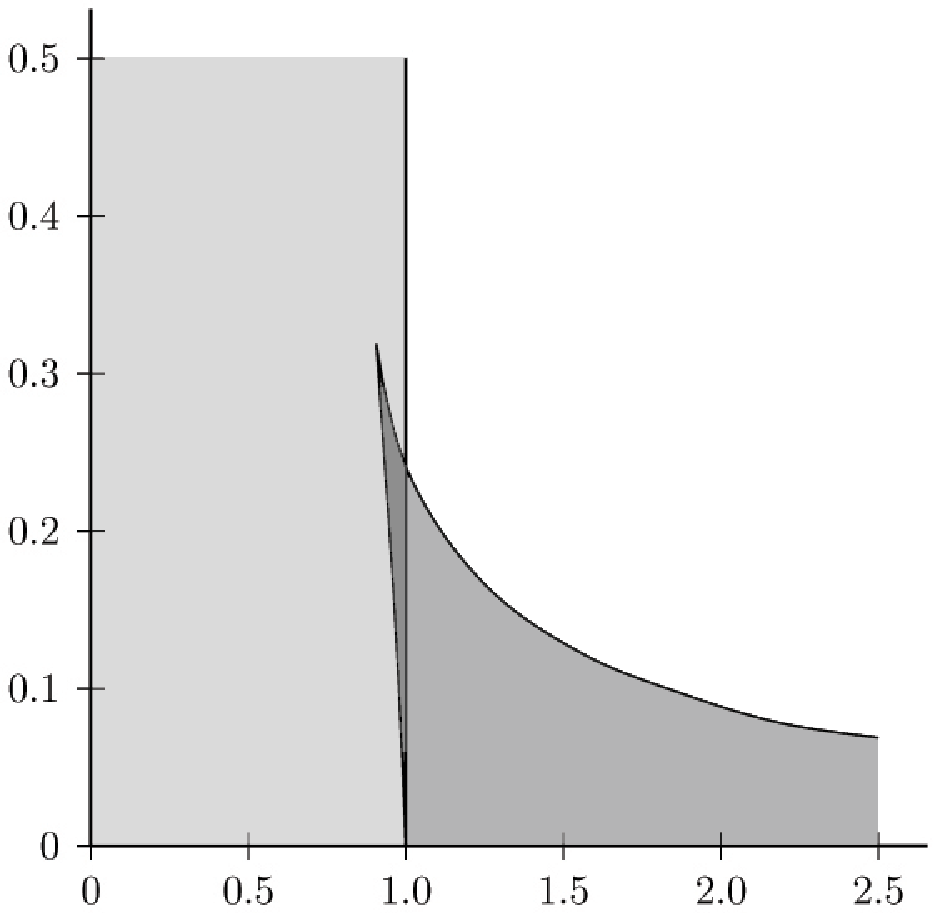}
} \\
\subfloat[][$\Lambda = 10^{-45} \text{km}^{-2}$]{
\includegraphics[width=0.31\textwidth]{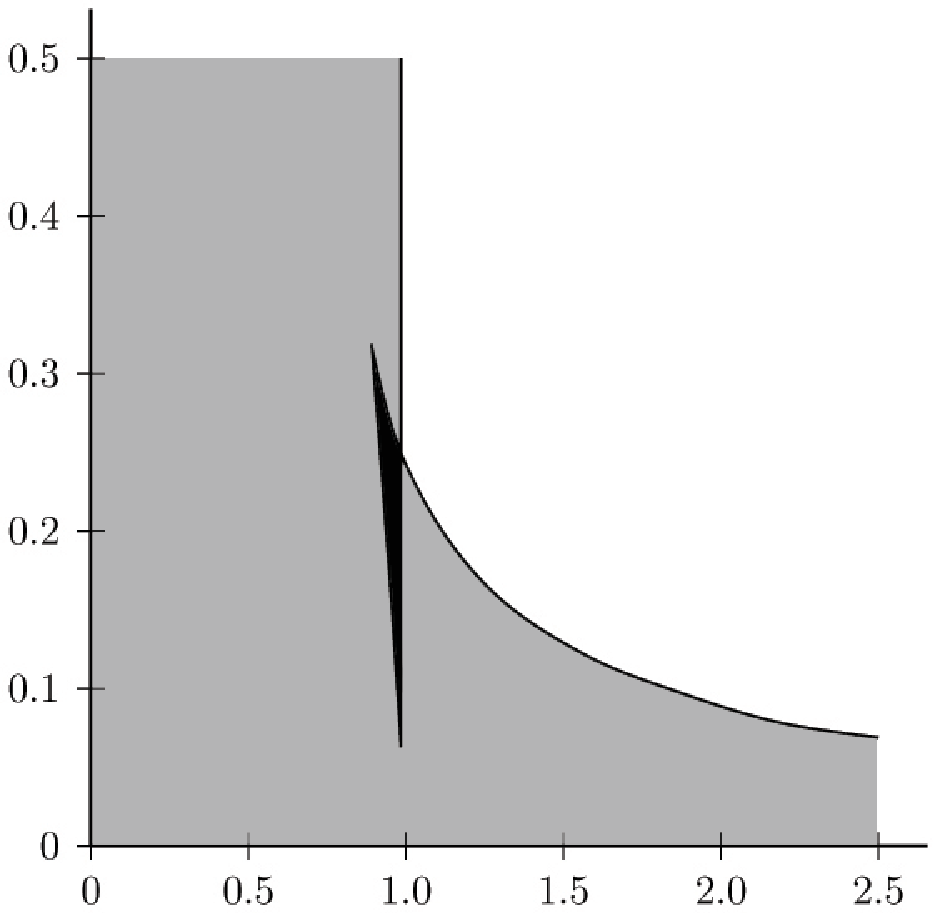}
}
\subfloat[][$\Lambda = 10^{-10} \text{km}^{-2}$]{
\includegraphics[width=0.31\textwidth]{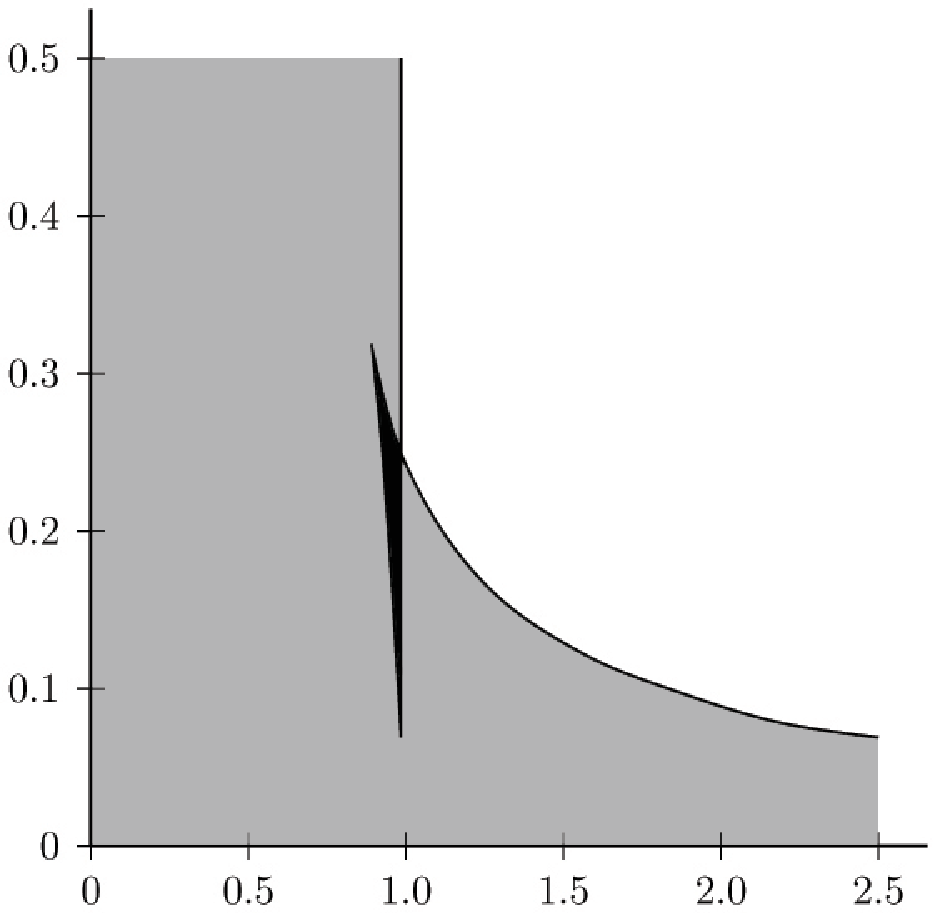}
}
\subfloat[][$\Lambda = 10^{-5} \text{km}^{-2}$]{
\includegraphics[width=0.31\textwidth]{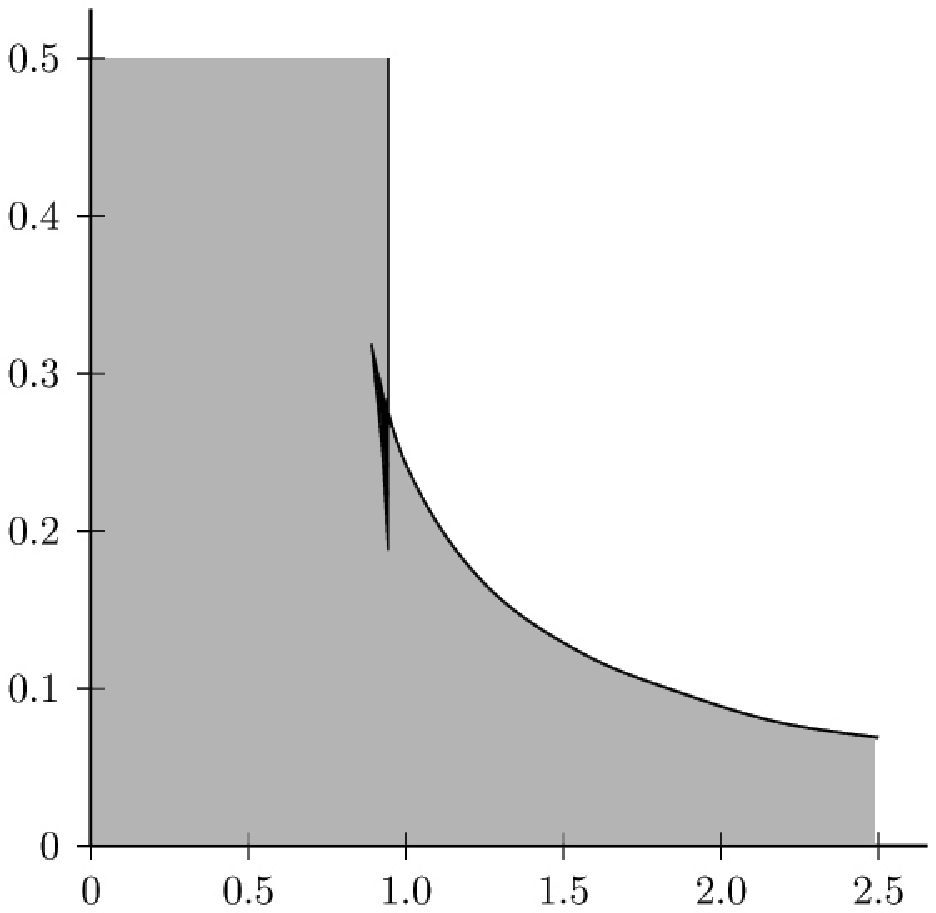}
}
\caption{The zeros of $P_5(u)$ in a $(\mu,\lambda)$--diagram for different values for the cosmological constant ($\mu$ is along the $x$--axis, $\lambda$ along the $y$--axis). The gray scales encode the numbers of positive real zeros of the polynomial $P_5$: black = 4, dark gray = 3, gray = 2, light gray = 1, white = 0. In the plot for $\Lambda = 0$ characteristic lines are shown (the left upper point of the dark gray region has the coordinates $(\mu = \frac{8}{9},\lambda=\frac{1}{3})$; the upper intersection point of the $\mu = 1$ line with the dark gray region is at $\lambda = \frac{1}{4}$).}
\label{fig:Dia}
\end{figure*}

Figure \ref{fig:Dia} shows the arrangement of zeros of $P_5(u)$ for some chosen values of $\rho$ as a $(\mu, \lambda)$ - diagram. The code of grayscales is as follows: black corresponds to four, dark gray to three, gray to two, light gray to one and white to no positive real zero. 

\subsection{Discussion with respect to $\Lambda$}

Based on Fig.~\ref{fig:Dia} we are now discussing the orbits related to different values of $\Lambda$. Each Plot in Fig.~\ref{fig:Dia} comprises all effective potentials (of the form shown in Fig.~\ref{Fig:SchwDeSitterEffPot}) for all possible values of $L$ and all particle energies $E$ and, thus, contains the complete  information about all orbits in Schwarzschild--(anti-)de Sitter space--times for a given value of $\Lambda$. 
\begin{enumerate}[a)]
\item Let us first consider $\Lambda = 0$. In this case the constant term in $P_5(u)$ vanishes and $P_5(u) = u^2 \tilde{P}_3(u)$ so that $u = 0$ is a zero of $P_5(u)$ with multiplicity $2$. The polynomial $\tilde{P}_3$ corresponds to the Schwarzschild space--time and has been extensively discussed in \cite{Hagihara31}. Nevertheless, let us examine some regions of $(\lambda,\mu)$ and possible orbits so that they can be directly compared with orbits for $\Lambda \neq 0$. As seen in Fig.~\ref{fig:Dia}(d), the straight line $\mu = 1$ divides the plot in two parts. For $\mu < 1$ there is an odd number of positive real zeros, i.e. we may not have any escape orbits in this regions. The light gray area corresponds to one real positive zero and, therefore, it is only a terminating orbit possible whereas in the dark gray region there may be in addition a bound orbit. For $\mu \geq 1$ there is an even number of positive real zeros and, thus, there is always an orbit which reaches infinity. The gray region corresponds to two real positive zeros and, hence, to a terminating and an escape orbit. The white region represents the case of no positive real zeros, i.e. an orbit which comes from infinity and falls into the singularity. Also, beside $u=0$ there is a further real zero with multiplicity $2$ on the straight line $\mu=1$.  
\item Let us compare now this with the case $\Lambda > 0$, see Fig.~\ref{fig:Dia}(e)-(g). We immediately recognize that left to the $\mu = 1$ line the plot significantly changed. In addition, we notice that for growing $\Lambda$ this straight line shifts a bit to the left. 
Left of the straight $\mu$--line there is now one more positive real zero in each region. This means that a particle which for $\Lambda = 0$ is in a light gray region now is in a gray region and, thus, may reach infinity. The same happens in the region which was dark gray for $\Lambda=0$ and is now black. A particle with $\mu<1$ in the small band now right of our straight line switched to a gray or white region depending on its $\lambda$ value. 

For a large positive cosmological constant the black area will disappear, that is, there will no longer be any bound orbit. This is clear from the following: First we introduce $\bar r := r/r_{\rm S}$. For $\Lambda = 0$ the effective potential $V_{\rm eff}$ possesses two different extrema $\bar r_{\pm}$ if $\lambda < \frac{1}{3}$. The smaller extrema $\bar r_-$ is a maximum whereas the larger $\bar r_+$ is a minimum. The extrema are bounded by $\bar r_- < 3$ and $\bar r_+ > \frac{2}{\lambda} - 3 > 3$. If we add the term containing $\Lambda > 0$, which is of the form of a parabola, $\bar r_-$ shifts to the left and $\bar r_+$ to the right. Thus, $\bar r_- < 3$ and $\bar r_+ > 3$ remain valid. In general, a second maximum $\bar r_\Lambda > \bar r_+$ will appear and $V_{\rm eff} \to - \infty$ for $r \to \infty$. It follows that there will be no bound orbit if the minimum $\bar r_+$ and, thus, the maximum $\bar r_\Lambda$ disappears. This is fulfilled if we choose $\Lambda$ such large that the gradient is negative for all $\bar r > 3$, for example $\Lambda r_S^2 > \frac{1}{18} - \frac{1}{54} \frac{1}{\lambda}$. Since $\frac{1}{\lambda} > 3$, the choice $\Lambda r_S^2 > \frac{1}{9}$ ensures that for any choice of $\lambda < \frac{1}{3}$ there will be no bound orbits. This is of course only a rough estimate which can be improved. 

\comment{
For a large positive cosmological constant the black area will disappear, that is, there will no longer be any bound orbit. This is clear from the following: First we introduce $\bar r := r/r_{\rm S}$. For $\Lambda = 0$ the effective potential possesses two extrema if $\lambda \geq 3$. For $\lambda = 3$ the two extrema coincide at the position $\bar r = 3$; for values of $\lambda > 3$ one maximum is located at $2 \lambda - \frac{3}{2} > \bar r > 2 \lambda - 3$, and the other is located $3 > \bar r > \frac{3}{2}$. The extremum at smaller $\bar r$ is a maximum while the second one at the larger $\bar r$ is a minimum.

Since for Schwarzschild $V_{\rm eff} \rightarrow 0$ for $\bar r \rightarrow \infty$ it is clear that in any case one can choose a $\Lambda > 0$ large enough so that the minimum disappears with the consequence that there will remain only at most two real positive zeros of $E = V_{\rm eff}$. Since the second derivative of the effective potential at the minimum is bounded it follows that there is a supremum $\Lambda_{\rm sup}$ over all $\lambda$ implying that for this $\Lambda_{\rm sup}$ the minimum disappears for all values of $\lambda$.
} 

\item If $\Lambda<0$, the situation changed the other way around. We again immediately see that the right side of the plot significantly changed. The straight $\mu$-line is no longer so easy to identify, but if we take into account the number of all real zeros, we can say that it shifts to the right when the absolute value of $\Lambda$ growths. An exception to this is the part for small $\lambda$. There the line bends to the right and allows a switch from the gray part for $\Lambda=0$ to the light gray part of $\Lambda<0$. Nevertheless, we can say that on the right side of the plot we have now an additional real positive zero and, thus, also an odd number of positive real zeros. This means that a particle is no longer able to reach infinity for any $(\lambda,\mu)$. In the region being white for $\Lambda = 0$ and which now is light gray we have now a bound terminating orbit. In the for $\Lambda = 0$ gray region which now is dark gray the escape orbit becomes bound.
\end{enumerate}

\begin{figure*}[t]
\subfloat[][$\Lambda =0$, $r_0=20.951 \text{km}$]{
\includegraphics[width=0.31\textwidth]{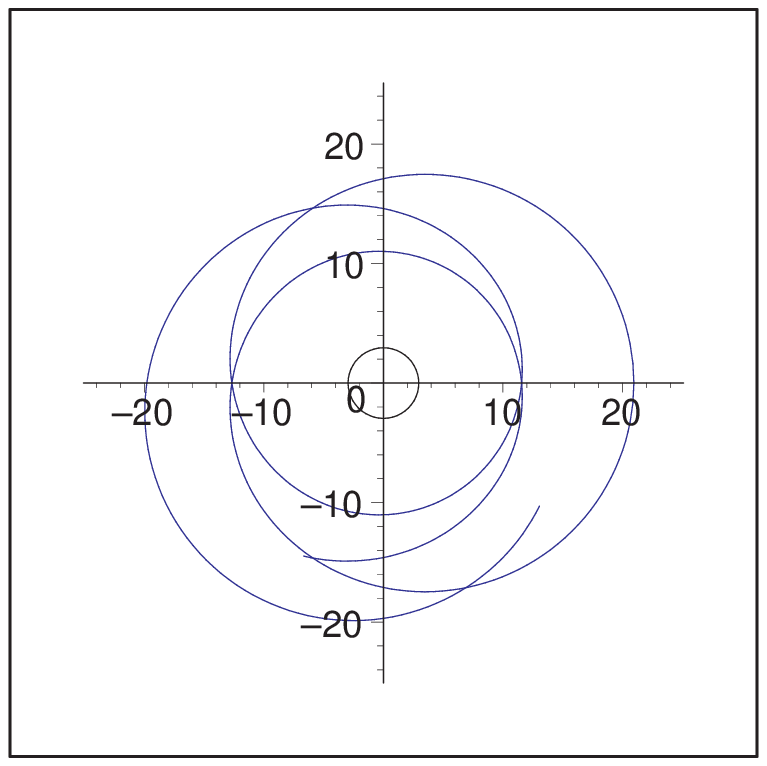}
}
\subfloat[][$\Lambda =0$, $r_0=5.010 \text{km}$]{
\includegraphics[width=0.31\textwidth]{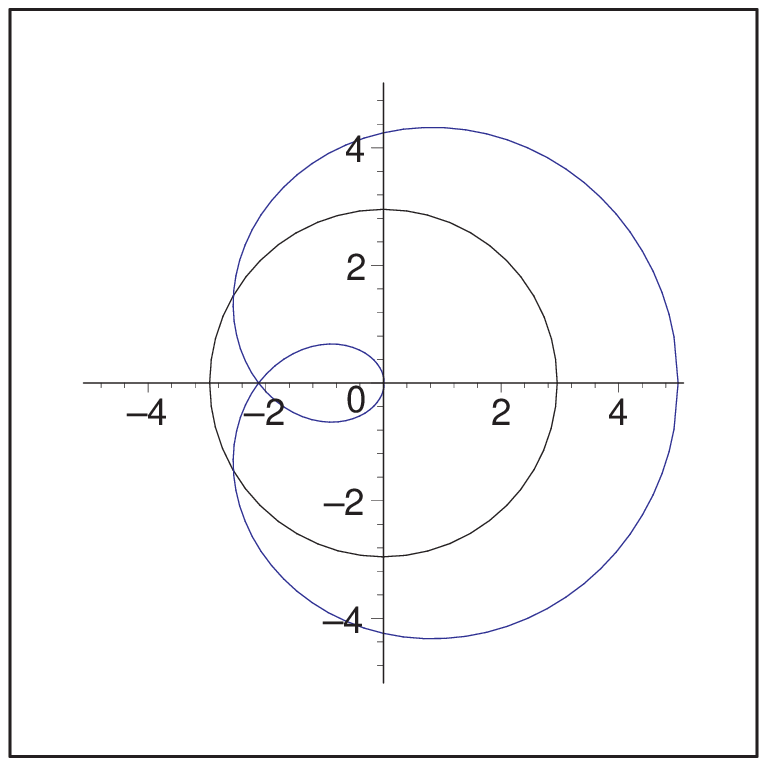}
} \\
\subfloat[][$\Lambda =10^{-5} \,\text{km}^{-2}$, $r_0=22.185 \text{km}$]{
\includegraphics[width=0.31\textwidth]{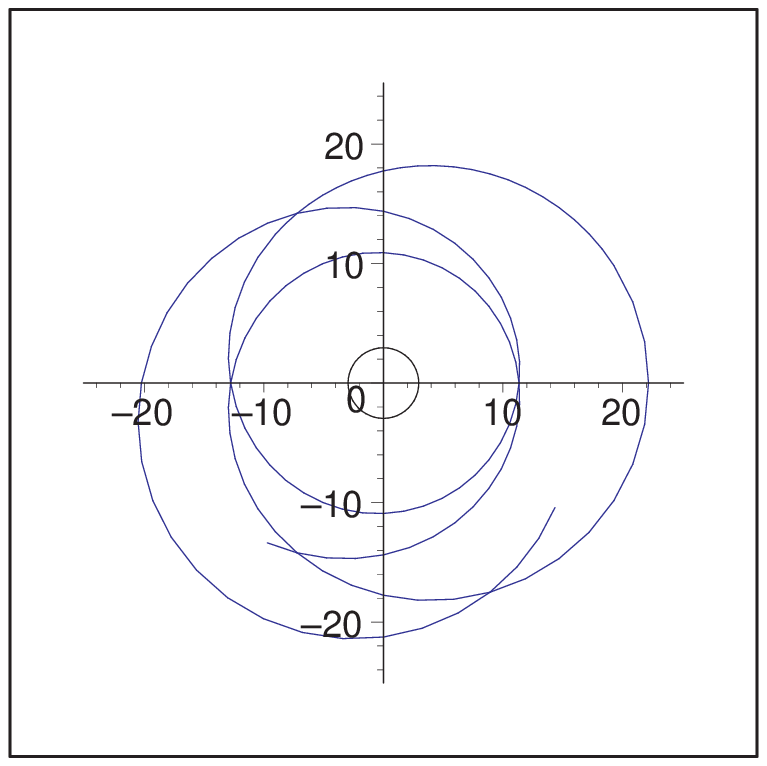}
}
\subfloat[][$\Lambda =10^{-5} \text{km}^{-2}$, $r_0=5.013 \text{km}$]{
\includegraphics[width=0.31\textwidth]{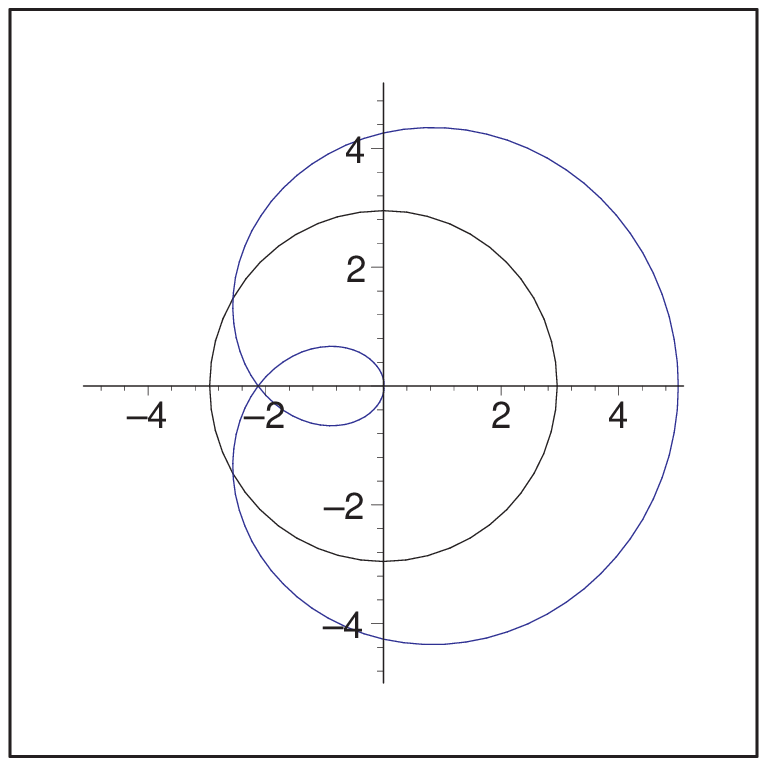}
}
\subfloat[][$\Lambda =10^{-5} \text{km}^{-2}$, $r_0=133.60 \text{km}$]{
\includegraphics[width=0.31\textwidth]{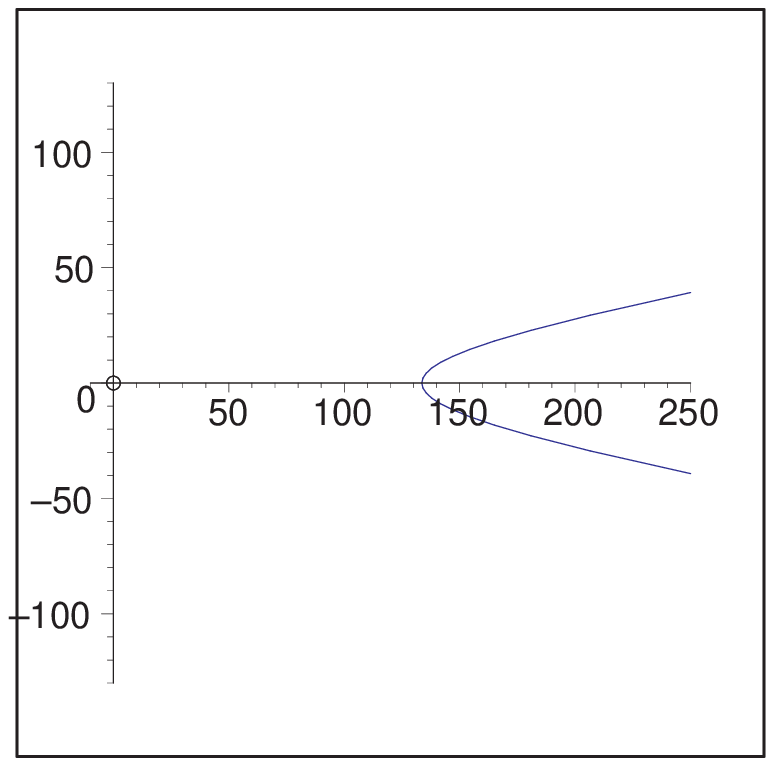}
}
\caption{Orbits for $\mu=0.92$ and $\lambda=0.28$. The upper row is for vanishing, the lower row for positive $\Lambda$. (a) and (c): bound orbits with perihelion shift. (b) and (d) terminating orbit ending in the singularity. (e): Reflection at the $\Lambda$--barrier. There is no analogue of (e) for $\Lambda = 0$. Black circles always indicate the Schwarzschild radius.  \label{Fig:Orbits1}}
\end{figure*}

\subsection{Plots of orbits}

\begin{figure*}[t]
\subfloat[][$\Lambda = 0$, $r_0=6.776 \text{km}$]{
\includegraphics[width=0.31\textwidth]{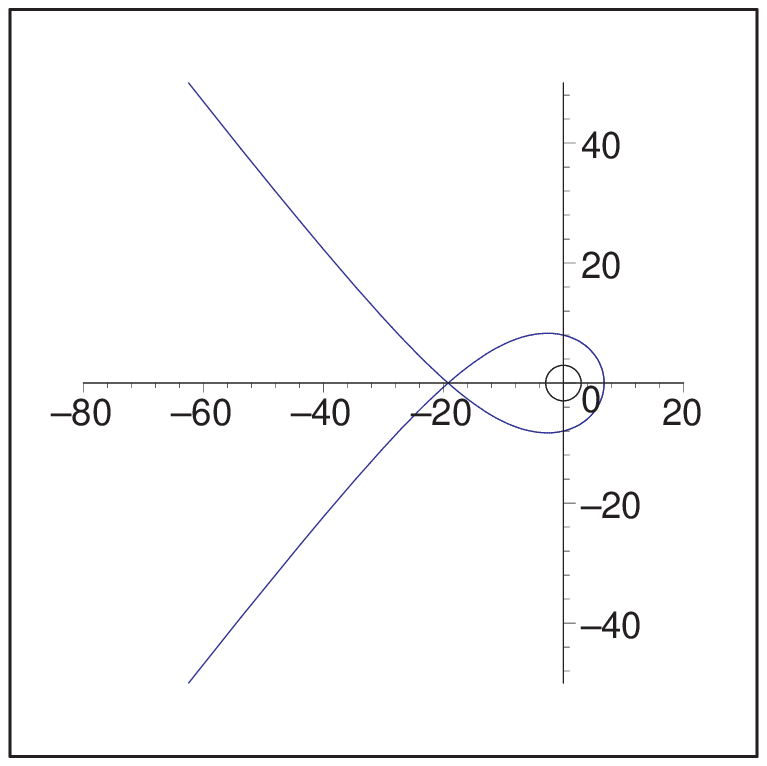}
}
\subfloat[][$\Lambda = 0$, $r_0=4.642 \text{km}$]{
\includegraphics[width=0.31\textwidth]{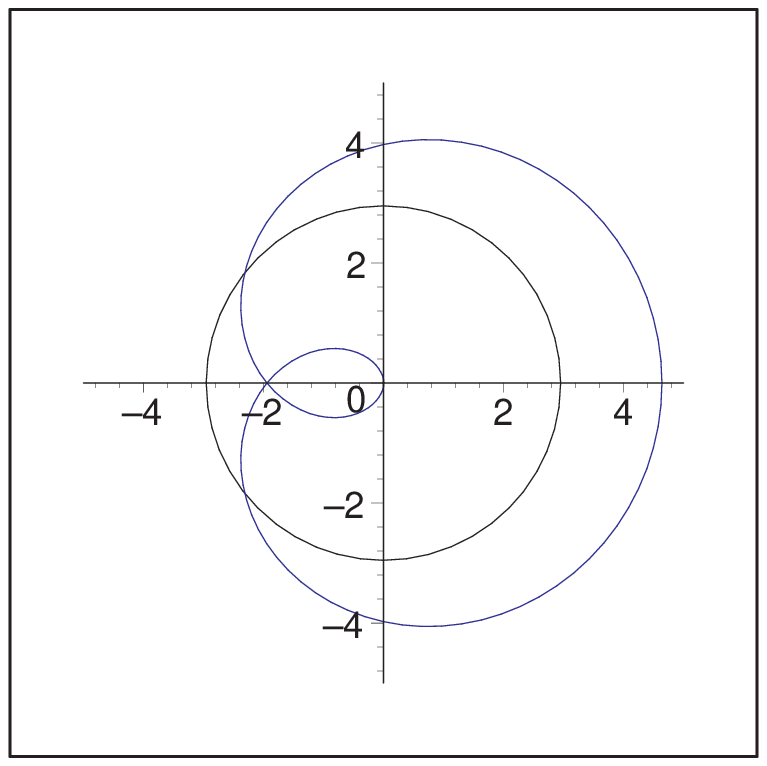}
} \\ 
\subfloat[][$\Lambda = -10^{-5} \text{km}^{-2}$, $r_0=185.37 \text{km}$]{
\includegraphics[width=0.31\textwidth]{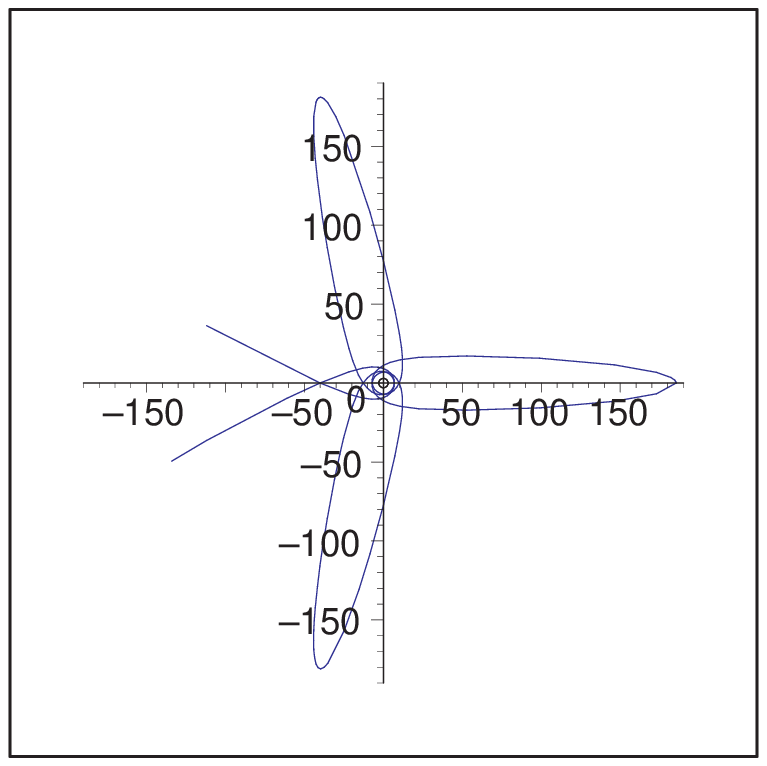}
}
\subfloat[][$\Lambda = -10^{-5} \text{km}^{-2}$, $r_0=4.639 \text{km}$]{
\includegraphics[width=0.31\textwidth]{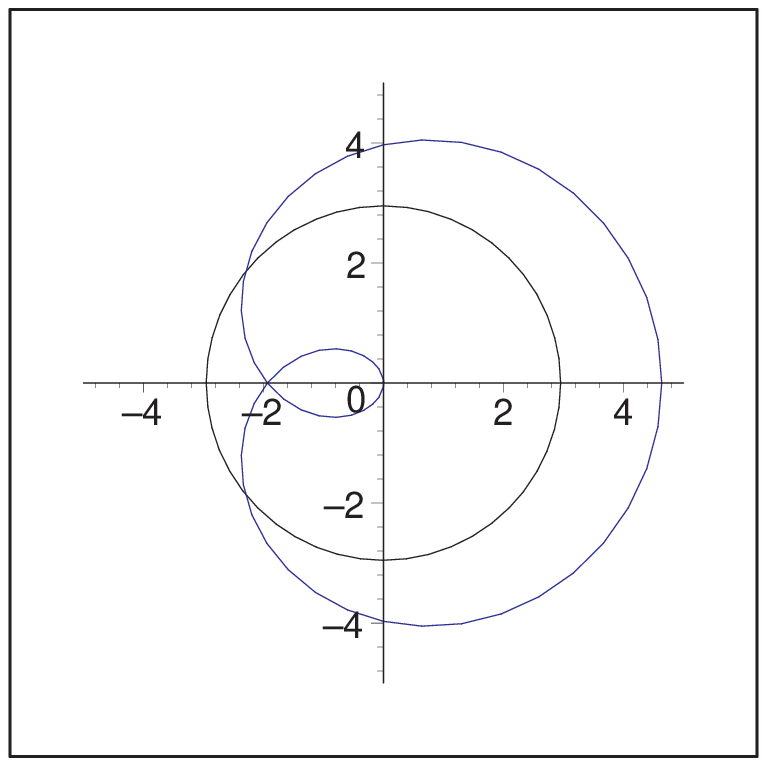}
}
\caption{Orbits for $\mu=1.1$ and $\lambda=0.2$. The upper row is for vanishing, the lower row for negative $\Lambda$. \label{Fig:Orbits2}}
\end{figure*}

\begin{figure*}[t]
\subfloat[][$\Lambda = 0$, $r_0=3.625 \text{km}$]{
\includegraphics[width=0.31\textwidth]{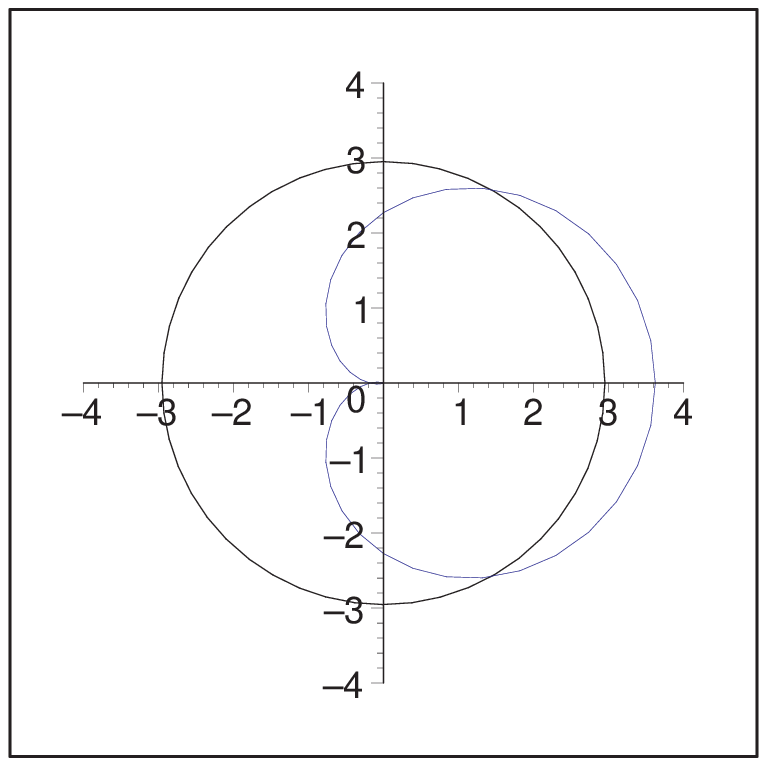}
}
\subfloat[][$\Lambda = 10^{-5} \text{km}^{-2}$, $r_0= 3.625 \text{km}$]{
\includegraphics[width=0.31\textwidth]{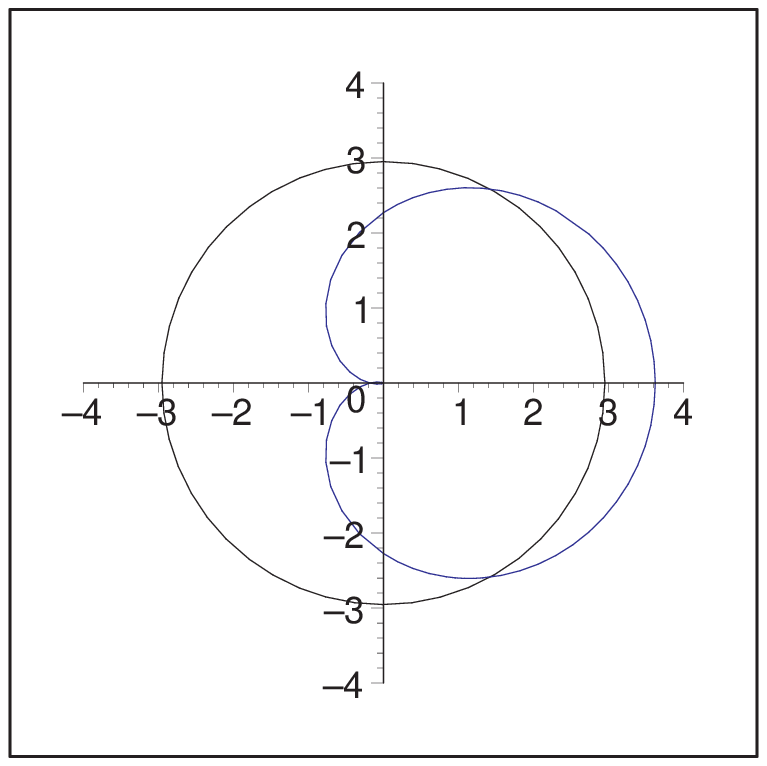}
}
\subfloat[][$\Lambda = 10^{-5} \text{km}^{-2}$, $r_0=237.61 \text{km}$]{
\includegraphics[width=0.31\textwidth]{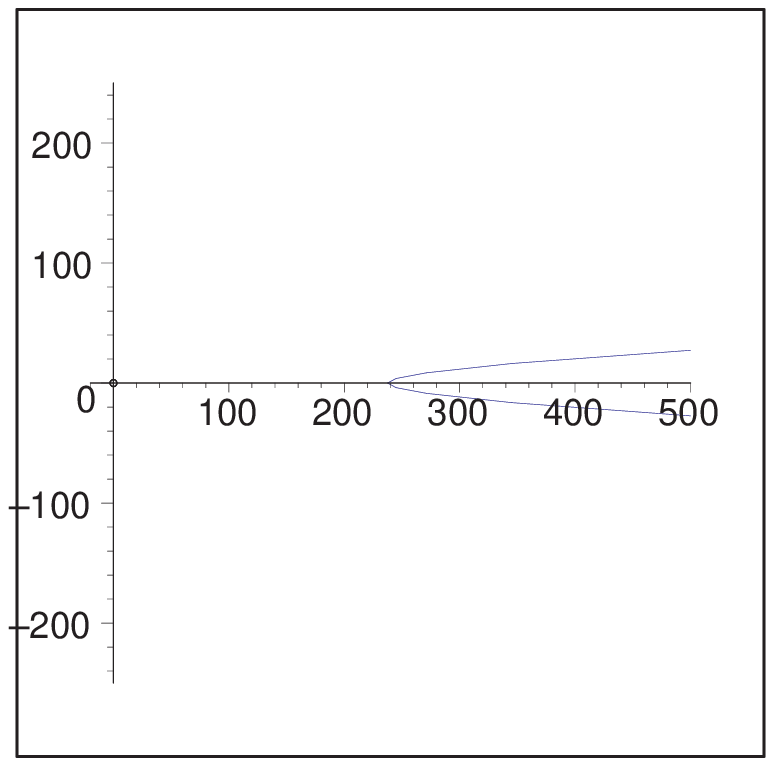}
}
\caption{Orbits for $\mu=0.8$ and $\lambda=0.2$. The upper graph is for vanishing, the lower graphs for positive $\Lambda$. There is no analogue of (c) for $\Lambda = 0$. \label{Fig:Orbits3}}
\end{figure*}

In Figs.~\ref{Fig:Orbits1}-\ref{Fig:Orbits3} some of the possible orbits are plotted. The figures are organized in order to highlight the influence of the cosmological constant. For all orbits in each figure the parameters $\mu$ and $\lambda$, that is, $E$ and $L$, are the same. The absolute value of the cosmological constant is chosen as $|\Lambda|=10^{-5}$ in all plots. All plots are created from the analytical solution derived in Sec.~\ref{solution}.

In Fig.~\ref{Fig:Orbits1} the parameters are $\mu = 0.92$ and $\lambda = 0.28$ which belong to the dark gray region in Fig.~\ref{fig:Dia}(d). For a vanishing cosmological constant this corresponds to a bound periodic orbit and to a bound terminating orbit ending in the singularity. The corresponding orbits are shown in Fig.~\ref{Fig:Orbits1}(a-b). For a positive cosmological constant the overall structure changes considerably since there will be a third type of orbits not present in the Schwarzschild case. Beside the terminating and bound orbits in Fig.~\ref{Fig:Orbits1}(c) which both look quite similar to the corresponding orbits in the Schwarzschild case there is an escape orbit which is repelled from the potential barrier related to the positive cosmological constant,  Fig.~\ref{Fig:Orbits1}(e).

The next parameter choice is $\mu=1.1$ and $\lambda=0.2$. For vanishing $\Lambda$ these parameters lay in the gray area of Fig.~\ref{fig:Dia}(d) denoting two zeros and, thus, correspond to a quasihyperbolic escape orbit,  Fig.~\ref{Fig:Orbits2}(a), and a terminating orbit ending in the singularity, Fig.~\ref{Fig:Orbits2}(b). For the chosen $\Lambda = - 10^{-5}\;{\rm km}^{-2}$ the situation changes dramatically as can be read off from Fig.~\ref{fig:Dia}(a) in comparison to Fig.~\ref{fig:Dia}(d): Now we have three zeros and, thus, one bound orbit and one terminating orbit ending in the singularity, see Figs.~\ref{Fig:Orbits2}(c-d). Switching on the negative cosmological constant makes an escape orbit a bound orbit. This of course has to be expected as one can see from Fig.~\ref{Fig:SchwDeSitterEffPot} that there are no escape orbits for negative cosmological constant. 

Our third choice of parameters is $\mu=0.8$ and $\lambda=0.2$. For $\Lambda = 0$, Fig.~\ref{fig:Dia}(d) this lays in the light gray region of one zero where is a terminating orbit only. This orbit is shown in Fig.~\ref{Fig:Orbits3}(a). For a positive cosmological constant $\Lambda > 0$ these parameters are in a gray region with two zeros indicating a terminating and an escape orbit, see Figs.~\ref{Fig:Orbits3}(b-c). The orbit in Fig.~\ref{Fig:Orbits3}(c) again is a reflection at the $\Lambda$--barrier. 

\section{On the Pioneer anomaly}

We apply the obtained analytical solution in order to decide whether a nonvanishing cosmological constant may have an observable influence on the Pioneer satellites. From \cite{NietoAnderson05} we may deduce the energy and angular momentum of the Pioneers after their last flybys at Jupiter and Saturn, respectively, with respect to the barycenter of the inner solar system, i.e. the Sun, Mercury, Venus and Earth-Moon. This means that we used the value
\begin{equation}
r_{\rm S} = \dfrac{2 GM}{c^2} = 2. \,953 \,266 \,762 \,363 \,45 \,\text{km} \,,
\end{equation} 
for the Schwarzschild radius, derived from $GM =  1. \,000 \,005 \,65 \,k^2 (\text{AU}^3/\text{day}^2)$ with Gauss' constant $k= 0. \,017 \,202 \,098 \,95$ defining the astronomical unit $\text{AU}$. Here all numbers are taken with 12 digits what corresponds to the today's accuracy of solar system ephemerides.

In the case of Pioneer 10, the velocity at infinity $v_{\infty} = 11.322 \,\text{km s}^{-1}$ taken from \cite{NietoAnderson05} gives us the energy per unit mass $E_M = c^2 + \dfrac{1}{2} v_{\infty}^2$ and therefore the parameter $\mu$,
\begin{equation}
\mu = \dfrac{E_M^2}{c^4} = 1. \,000 \,000 \,001 \,43\,.
\end{equation}
The angular momentum per unit mass is given by $L_M = q \, v$, where $v = \sqrt{2 GM \left( \frac{1}{q} + \frac{1}{2a} \right)}$ is the velocity at periapsis distance $q$; $a$ is the semimajor axis. From this we derive the parameter $\lambda$,
\begin{equation}
\lambda = \dfrac{r_{\rm S}^2 c^2}{L_M^2} = 2.  \,855 \,572 \,373 \,82 \cdot 10^{-9}\,.
\end{equation}
In the case of Pioneer 11 we obtain for the parameters $\mu$ and $\lambda$
\begin{equation}
\begin{split}
\mu & = 1. \,000 \,000 \,001 \,22 \,,\\
\lambda & = 1. \,340 \,740 \,574 \,59 \cdot 10^{-9}\,.
\end{split}
\end{equation}

With these coefficients we now can determine the exact orbits of Pioneer 10 and 11 in the cases $\Lambda=0$ and $\Lambda=10^{-45} \text{km}^{-2}$. From these exact orbits we calculated the differences in position (in m) for a given angle $\varphi$ (in rad) and the difference in the angle (in rad) for a given distance $r$ (in m) of a test particle moving in a space--time with and without cosmological constant. The Pioneer anomaly appeared in a heliocentric distance from about 20 to 70 AU. For $r$ in this range, we compute now the difference $\varphi_{\Lambda=0}(r) - \varphi_{\Lambda \neq 0}(r)$ in azimuthal position with and without cosmological constant for both craft. Regarding Pioneer 10, the difference is in the scale of $10^{-19}$rad, which corresponds to an azimuthal difference in position of about $10^{-6}$m. For Pioneer 11, the difference is in the scale $10^{-18}$rad, which corresponds to an azimuthal difference in position of about $10^{-5}$m.

The range of 20 to 70 AU corresponds to an angle between $0.4 \pi$ and $0.6 \pi$ if $\varphi_0=0$ corresponds to the periapsis. In this range, we compute the radial difference $r_{\Lambda=0}(\varphi) - r_{\Lambda\neq 0}(\varphi)$ also for both craft. For Pioneer 10 we obtain a difference in the scale of $10^{-5}$m, for Pioneer 11 in the scale of $10^{-4}$m.

Therefore we can say, that for the present value of the cosmological constant the form of the Pioneer 10 orbit practically does not change. For a definite estimate of the differences of the Pioneer orbits in Schwarzschild and Schwarzschild--de Sitter space--time one of course has to analyze the time course of these orbits. However, the time variable is influenced by the cosmological constant in the same way as the radial coordinate so that no change in our statement will occur. Therefore, the influence of the cosmological constant on the orbits cannot be held responsible for the observed anomalous acceleration of the Pioneer spacecraft. 

\section{Periastron advance of bound orbits}

In the case that $P_5$ has at least three real and positive zeros, we may have a bound orbit for some initial values. The periastron advance $\Delta_{\text{peri}}$ for such a bound orbit is given by the difference of the $2\pi$--periodicity of the angle $\varphi$ and the periodicity of the solution $r(\varphi)$ (which is the same as the periodicity of $u(\varphi)$). Let us assume that the bound orbit corresponds to the interval $[e_k,e_{k+1}]$, where $e_k$ and $e_{k+1}$ are real and positive zeros of $P_5$, and that the path $a_i$ surrounds this real interval. Then the periastron advance is given by 
\begin{equation}\label{perihelshift}
\Delta_{\text{peri}} = 2\pi - 2\omega_{2i} = 2\pi - 2 \int_{e_k}^{e_{k+1}} \frac{x dx}{\sqrt{P_5(x)}}\,,
\end{equation}
where $2 \omega_{2i}$ is an element of the (canonically chosen) $2 \times 4$ matrix of periods $(2 \omega, 2\omega')$ of $\sqrt{P_5}$, see Eq.~\eqref{periodmatrices}. We now calculate the post--Schwarzschild limit of this periastron advance in the case that the considered bound orbit is also bound in Schwarzschild space--time.

For doing so we first expand $x/\sqrt{P_5(x)}$ to first order in $\Lambda$
\begin{equation}\label{taylor}
\frac{x}{\sqrt{P_5(x)}} \approx \frac{1}{\sqrt{P_3(x)}} - \frac{1}{6} r_{\rm S}^2 \frac{x^2+\lambda}{x^2 P_3(x) \sqrt{P_3(x)}} \Lambda \, ,
\end{equation}
where $P_3(x) = x^3-x^2+\lambda x+ \lambda(\mu-1)$ is the polynomial for the corresponding Schwarzschild case given by $\Lambda = 0$.
  
\begin{figure}[t]
\begin{center}
\includegraphics[width=0.5\textwidth]{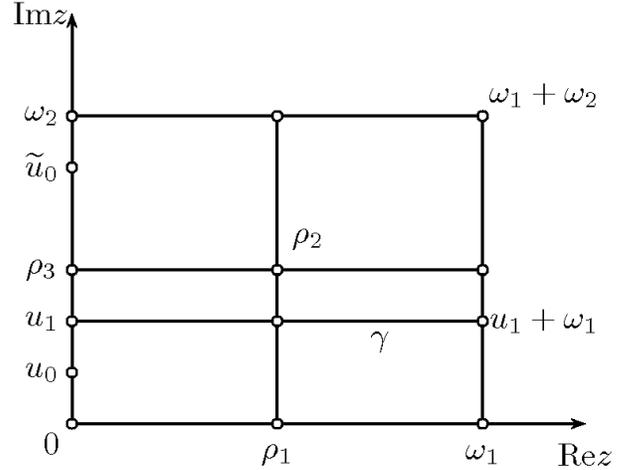}
\end{center}
\caption{The fundamental rectangle $R$}
\label{fig:rectangle}
\end{figure}

In the next step we have to integrate both terms involving $P_3$ within the Weierstrass formalism, see for example \cite{Hurwitz64}. Employing the substitution $x = 4z + 1/3$ we rewrite $P_3$ in a Weierstrass form
\begin{equation}
P_3(x) = 4^2 ( 4z^3-g_2z-g_3 ) = 4^2 P_W(z)\,,
\end{equation}
where
\begin{align}
g_2 & = \frac{1}{12} - \frac{1}{4} \lambda \\
g_3 & = \frac{1}{16} \left( \frac{2}{27} + \frac{2}{3} \lambda - \lambda \mu \right) 
\end{align}
are the Weierstrass invariants. We assume that the orbit under consideration is bound not only in the Schwarzschild--de Sitter but also in the corresponding Schwarzschild space--time. This means that the three largest real zeros of $P_5$ are positive and, thus, the zeros $z_1 > z_2 > z_3 > - \frac{1}{12}$ of $P_W$ are all real. The square root $\sqrt{P_W}$ is branched over $z_1, z_2$ and $z_3$ and, thus, the elliptic function $\wp$ based on $\sqrt{P_W}$ has a purely real period $\omega_1$ and a purely imaginary period $\omega_2$. They are given by
\begin{equation}\label{periods_Weier}
\begin{split}
\omega_1 & = \oint_{A} \frac{dz}{\sqrt{P_W(z)}} \\
\omega_2 & = \oint_{B} \frac{dz}{\sqrt{P_W(z)}}
\end{split}
\end{equation}
where the path $A$ runs around the branch cut from $z_3$ to $z_2$ and the path $B$ around $z_2$ and $z_1$, both clockwise. The branch of the square root in \eqref{periods_Weier} is chosen such that $\sqrt{P_W} > 0$ on $[z_3,z_2]$ and, thus, $\sqrt{P_W}$ negatively imaginary on $[z_2,z_1]$. The branch points of $\wp$ can be expressed in terms of the periods: $z_1 = \wp(\rho_1)$, $z_2 = \wp(\rho_2)$ and $z_3 = \wp(\rho_3)$ with $\rho_1 = \omega_1/2$, $\rho_2 = (\omega_1+\omega_2)/2$ and $\rho_3 = \omega_2/2$. The fundamental rectangle in the complex plane spanned by the periods $\omega_1, \omega_2$ of $\wp$ is denoted by $R = \{ x\omega_1 + y\omega_2 \,|\, 0 \leq x,y < 1 \}$, see Fig.~\ref{fig:rectangle}. 

Let the three biggest real and positive zeros of $P_5$ be given by $x_1 > x_2 > x_3 > 0$. Then, for the canonical choice of the matrix of periods $\omega$ of $\sqrt{P_5}$, the integration path $a_i$ runs from $x_3$ to $x_2$ and back with conversed sign of the square root. Let the path $\gamma$ be the preimage of $a_i$ by $u \mapsto \wp(u) = z$ in the fundamental rectangle $R$. For a positive cosmological constant $\Lambda$, we have $x_3 < z_3 < z_2 < x_2$ and, thus, $\gamma$ starts at some purely imaginary $\gamma(0) = u_1 \in R$ with $0< \Im(u_1) \leq \Im(\rho_3)$ and goes straight to $\gamma(1) = u_1+\omega_1$. Then, for any rational function $F$, we obtain
\begin{equation}
\oint_{a_i} F(z) \frac{dz}{\sqrt{P_W(z)}} = \int_\gamma F(\wp(u)) du \,.
\end{equation}
This is derived from the differential equation
\begin{equation}\label{dgl_Weier}
\wp'(u) = \sqrt{4\wp(u)^3 - g_2 \wp(u) - g_3} = \sqrt{P_W(\wp(u))}\,,
\end{equation}
where the branch of the square root was chosen to be consistent with the sign of $\wp'$.

The integration of the first part on the right-hand side of \eqref{taylor} is straightforward and yields the Schwarzschild period
\begin{equation}\label{firstpart}
\oint_{a_i} \frac{dx}{\sqrt{P_3(x)}} = \oint_{a_i}  \frac{dz}{\sqrt{P_W(z)}} = \int_{u_1}^{u_1+\omega_1} du =  \omega_1\,.
\end{equation}
The integration of the second part on the right-hand side of \eqref{taylor} is more involved and is performed in Appendix B. As a result we obtain the first order approximation of the periastron shift with respect to $\Lambda$:
\begin{align}
& \Delta_{\text{peri}} = 2\pi - \oint_{a_i} \frac{x dx}{\sqrt{P_5(x)}} \nonumber \\
& = 2 \pi - \Bigg\{ \omega_1 + \Lambda \, \frac{r_{\rm S}^2}{96} \Bigg[\sum_{j=1}^3 \frac{\eta_1 + z_j \omega_1}{\wp''(\rho_j)^2} \left( 1+ \frac{\lambda}{\left( 4z_j+\frac{1}{3} \right)^2} \right)\nonumber \\
& \qquad + \lambda \left( \frac{2\eta_1 - \frac{1}{6} \omega_1}{16 \wp'(u_0)^{4}} + \frac{6}{16} \frac{\wp''(u_0)}{\wp'(u_0)^5} ( - \eta_1 u_0 + \zeta(u_0)) \right) \Bigg] \Bigg\}\nonumber\\
& \quad + {\cal O}(\Lambda^2) \,, \label{postSchwarzschildPeriastron}
\end{align}
where $u_0$ is such that $\wp(u_0) = - \frac{1}{12}$.

The terms in this expression involving $\rho_j$ and $u_0$ can partly be replaced by terms containing the Weierstrass invariants $g_2$ and $g_3$. From the differential equation \eqref{dgl_Weier} we derive
\begin{align}
\wp'(u_0) = \sqrt{4\wp(u_0)^3 - g_2\wp(u_0) - g_3} = \sqrt{- \frac{1}{432} + \frac{g_2}{12} - g_3}\,.
\end{align}
The first derivative of \eqref{dgl_Weier} yields $2 \wp'' = 12 \wp^2 - g_2$ and, thus, gives
\begin{align}
\wp''(\rho_j) = 6 z_j^2 - \frac{1}{2} g_2 \quad \text{and} \quad \wp''(u_0) = \frac{1}{24} - \frac{1}{2} g_2 \, ,
\end{align}
where the $g_2$, $g_3$, as well as the zeros of $P_W$ can be expressed by $\mu$ and $\lambda$. 

The result \eqref{postSchwarzschildPeriastron} gives the post--Schwarzschild periastron shift in a closed algebraic form. The advantage of this result is that no further integration is needed. Another advantage lies in the fact that only elliptic functions and related quantities are used which are well described and tabulated in mathematical books and which are also well implemented in common commercial math programs. What is still left to do is to express the result \eqref{postSchwarzschildPeriastron} in terms of, e.g., $r_{\rm min}$ and $r_{\rm max}$ or, equivalently, in terms of the semimajor axis and the eccentricity. These quantities are directly observable and also have the advantage that an expansion in terms of $m/r_{\rm min}$ and $m/r_{\rm max}$ can be performed giving in addition a post--Newtonian expansion. This will be described elsewhere. 

Let us apply these formulas to the perihelion advance of Mercury and compare with the results of Kraniotis and Whitehouse, \cite{KraniotisWhitehouse03}. 
We take the values $r_S = 2953.250 08 \rm m$ for the Schwarzschild-radius, $\dfrac{r_S}{L_M^2} = 1.184 962 712 826 8641 \times 10^{-24} \,{\rm m^2}/{\rm s^2}$ for the angular momentum per unit mass $L_M$ and $\sqrt{E_M} = 0.029 979 245 417 779 875 \times 10^{10} \,{\rm m}/{\rm s}$ for the energy per unit mass $E_M$ given in \cite{KraniotisWhitehouse03}. These values lead to the zeros
\begin{align*}
z_1 & = 0.166 666 640 041 880\,,\\
z_2 & = - 0.083 333 317 283 501\,,\\
z_3 & = - 0.083 333 322 758 379
\end{align*}
of $P_W$ and to the periods
\begin{align*}
\omega_1 & = 3.141 592 904 522 524 6\,,\\
\omega_2 & = 20.409 391 639 385 179 9 i\,,\\
\tau & = 6.496 510 610 908 418 7 i \,,
\end{align*}
which all compare well to the results in \cite{KraniotisWhitehouse03}. Also the physical data, i.e. the aphel $r_A$, the perihel $r_P$ and the perihelion advance in Schwarzschild-space-time $\Delta^{\rm S}$, compare well to \cite{KraniotisWhitehouse03} and also to observations \footnote{http://history.nasa.gov/SP-423/intro.htm.}:
\begin{align}
r_A & = 6.981 708 938 652 731 \cdot 10^{10} \rm m\,,\nonumber\\
r_P & = 4.600 126 052 898 539 \cdot 10^{10} \rm m\,, \\
\Delta^{\rm S} & = 42.980 165 \, \rm arcsec \, \rm cy^{-1}\,. \nonumber
\end{align}
Here we used the rotation period $87.97 \, \rm days$ of Mercury and 100 SI-years per century to determine the unit $\rm arcsec \, \rm cy^{-1}$.

The first order post-Schwarzschild correction $\Delta^{\rm SdS}_{\rm corr}$ to the perihelion advance can now be calculated from formula \eqref{postSchwarzschildPeriastron}. For a cosmological constant of $\Lambda = 10^{-51} \rm m^{-2}$ we obtain for the parameters which appear in the expansion \eqref{postSchwarzschildPeriastron}
\begin{align*}
\eta_1 & = -.261 799 370 013 130 8 \,,\\
\wp'^2(u_0) & = -1.697 262 234 791 570 8 \cdot 10^{-16}\,, \\
\wp''(u_0) & = 1.331 239 218 453 965 7 \cdot 10^{-8}\,,\\
\zeta(u_0) & = 1.012 109 319 614 658 4 \,i \,.
\end{align*}
This leads to a correction of 
\begin{equation}
\Delta^{\rm SdS}_{\rm corr} = 5.82 \cdot 10^{-17} \, \rm arcsec \, \rm cy^{-1} \,.
\end{equation}
This result also compares well to \cite{KraniotisWhitehouse03} where the perihel advance of Merkur does not change within the given accuracy when considered in Schwarzschild-de Sitter space-time. The value of the correction is also far beyond the measurement accuracy of $0.002 \, \rm arcsec \, \rm cy^{-1}$ for the perihelion advance of Mercury.

However, for an extreme case the influence of the cosmological constant on the periastron advance may become measurable. The orbital data of quasar QJ287 reported in \cite{Valtonen08, Valtonenetal08} indicates that the correction to the periastron advance $\Delta^{\rm SdS}_{\rm corr}$ will be some orders of magnitude larger than the correction in the case of Merkur. Indeed, when we calculate from this data the energy parameter $\mu$ and the angular momentum parameter $\lambda$,
\begin{equation}
\mu = 0.982 166 \,, \quad \lambda = 0.092 317\,,
\end{equation}
we obtain 
\begin{equation}
\Delta^{\rm SdS}_{\rm corr} \approx 10^{-13} \, \rm arcsec \, \rm cy^{-1} \,.
\end{equation}

\section{Geodesics in higher dimensional Schwarzschild space--times}

We want to show here that our method for solving the equation of motion in Schwarzschild--(anti-)de Sitter space--times can also be applied to solve the geodesic equation in, e.g., higher dimensional Schwarzschild space--times. The metric of a Schwarzschild space--time in $d$ dimensions is given by \cite{Tangherlini63}
\begin{align}
ds^2 & = \left(1 - \left(\frac{r_{\rm S}}{r}\right)^{d-3} \right) dt^2 - \left(1 - \left(\frac{r_{\rm S}}{r}\right)^{d-3} \right)^{-1} dr^2\nonumber\\
& \quad - r^2 d\Omega_{d-2}^2 \,,
\end{align}
where $d\Omega^2_1 = d\varphi^2$ and $d\Omega_{i+1}^2 = d\theta_i + \sin^2 \theta_i d\Omega_i^2$ for $i \geq 1$. Because of spherical symmetry, we again restrict the considerations to the equatorial plane by setting $\theta_i = \frac{\pi}{2}$ for all $i$. With the conserved energy $E$ and angular momentum $L$ as well as the substitution $u=\frac{r_{\rm S}}{r}$ the geodesic equation reduces to
\begin{equation}\label{Schwdim}
\left(\frac{du}{d\varphi}\right)^2 = u^{d-1} + \lambda u^{d-3} - u^2 + \lambda(\mu-1) = P_{d-1}(u)\,,
\end{equation}
where the parameters $\lambda=\frac{r_{\rm S}^2}{L^2}$ and $\mu=E^2$ have the same meaning as in the Schwarzschild--(anti-)de Sitter case \eqref{parameter}. For $d=4$ this equation reduces of course to the Schwarzschild case \cite{Hagihara31}. For $d=5$ a substitution $u=\frac{1}{x}+l$ where $l$ is a zero of $P_4$ reduces the differential equation \eqref{Schwdim} to 
\begin{equation}
\left(\frac{dx}{d\varphi}\right)^2 = b_3 x^3 + \ldots + b_0 x^0\,.
\end{equation}
With an additional substitution $x=\frac{1}{b_3} \left( 4y - \frac{b_2}{3} \right)$ this equation acquires the form \eqref{Dgl_ell} which can be solved by Weierstrass' elliptic functions. 

In the case of a $d=6$--dimensional Schwarzschild space--time, however, the differential equation \eqref{Schwdim} comprises a polynomial of degree five on the right-hand side. This now can be solved by means of our method. The only difference is that the physical angle $\varphi$ is now given by
\begin{equation}
\varphi - \varphi_0 = \int_{u_0}^u \frac{du'}{\sqrt{P_5(u')}} = \int_{u_0}^u dz_1
\end{equation}
what corresponds to $dz_1$ rather than to $dz_2$ as it was in the Schwarzschild--(anti-)de Sitter case. This means that the solution of the geodesic equation in six--dimensional Schwarzschild space--time is given by
\begin{equation}
r(\varphi) = \frac{r_{\rm S}}{u(\varphi)} = - r_{\rm S} \frac{\sigma_2(\varphi_{\vec \Theta,6})}{\sigma_1(\varphi_{\vec \Theta,6})}\,,
\end{equation}
where $\vec \varphi_{\Theta,6} = \vek{\varphi-\varphi'_0}{\varphi_1}$. Here $\varphi_1$ is chosen in such a way that $(2 \omega)^{-1} \vec \varphi_{\Theta,6}$ is an element of the theta divisor $\Theta_{\vec K_\infty}$ and $\varphi'_0 = \varphi_0 + \int_{u_0}^\infty dz_1$ depends only on the initial values $u_0$ and $\varphi_0$.

The case $d=7$ corresponds to a polynomial $P_6$ of degree six. If we apply a substitution $u=\frac{1}{x} + l$ where $l$ is a zero of $P_6$, we obtain the differential equation
\begin{equation}
\left(x \frac{dx}{d\varphi}\right)^2 = b_5 x^5 + \ldots b_0 x^0
\end{equation}
with some constants $b_i$. This can be solved in exactly the same way as the differential equation \eqref{Dgl_order5}. The solution is
\begin{equation}
r(\varphi) = \frac{r_{\rm S}}{u(\varphi)} = - r_{\rm S} \frac{\sigma_2(\varphi_{\vec \Theta,7})}{\sigma_1(\varphi_{\vec \Theta,7})}\,,
\end{equation}
where $\varphi_{\vec \Theta,7} = \vek{\varphi_1}{\varphi - \varphi'_0}$ and, again, $\varphi_1$ is selected such that $(2 \omega)^{-1} \varphi_{\vec \Theta,7}$ is an element of the theta divisor $\Theta_{\vec K_\infty}$ and $\varphi'_0 = \varphi_0 + \int_{u_0}^\infty dz_2$. The only difference to the solution of the geodesic equation in Schwarzschild--(anti-)de Sitter space--time is that the periods $\omega$ and $\omega'$ and, hence, the matrix $\tau$ will be different due to the different coefficients in the polynomials appearing on the right-hand side of each differential equation.

\begin{figure}[t]
\centering
\includegraphics[width = 0.31 \textwidth]{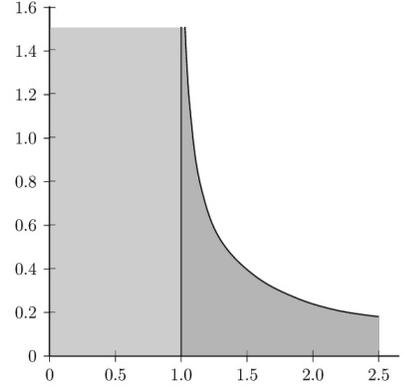}
\caption{Arrangement of zeros of the polynomial $P_{d-1}$ for $d=6$. The gray scale code is the same as in Fig.~\ref{fig:Dia}.}
\label{fig:6dDia}
\end{figure}

\begin{figure*}[t]
\subfloat[][Terminating orbit for $\mu=1.1$, $\lambda=0.3$ (gray region in Fig.~\ref{fig:6dDia}).]{
\includegraphics[width=0.31\textwidth]{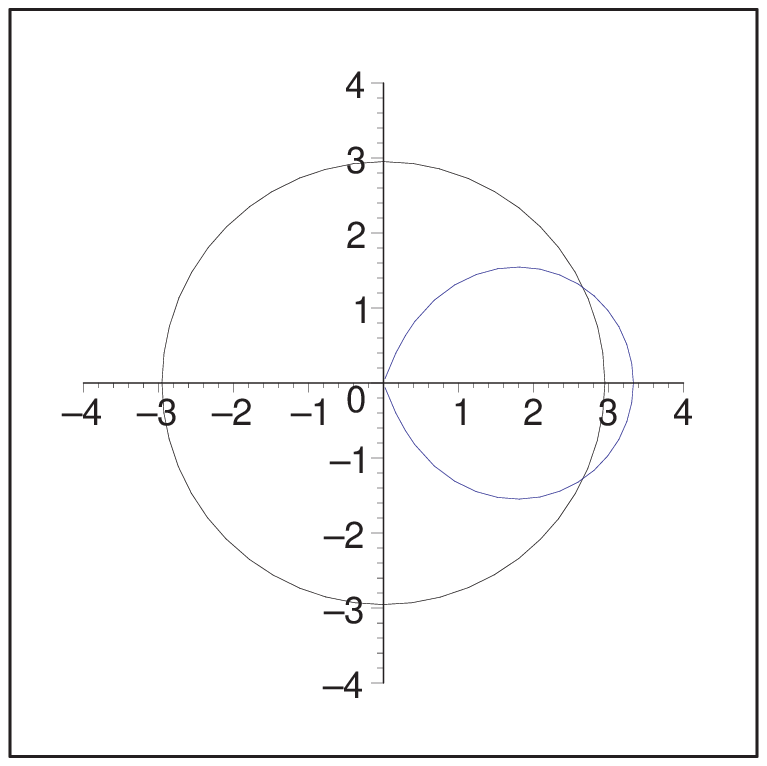}
}
\subfloat[][Escape orbit for $\mu=1.1$, $\lambda=0.3$ (gray region in Fig.~\ref{fig:6dDia}).]{
\includegraphics[width=0.31\textwidth]{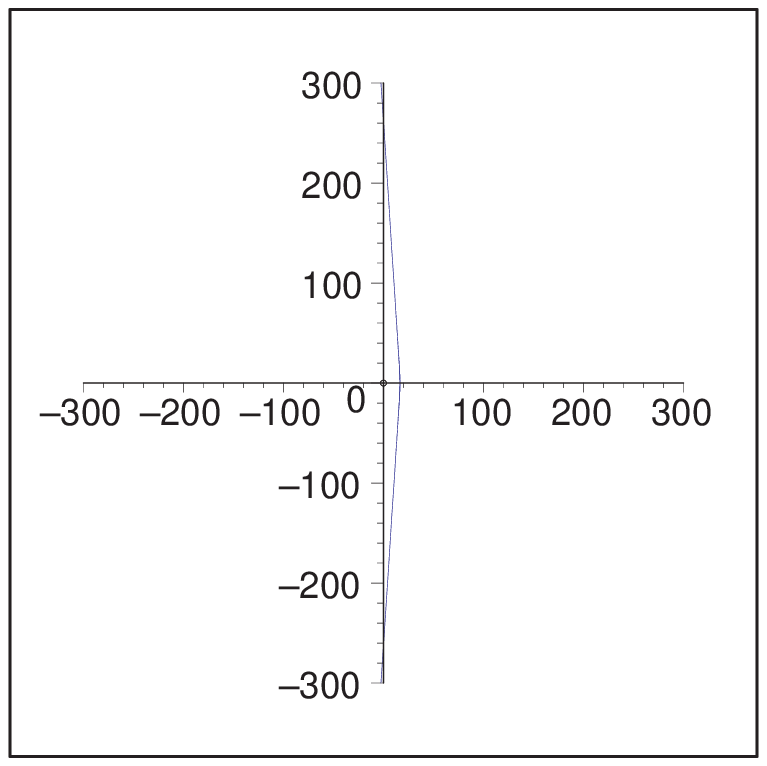}
} \\ 
\subfloat[][Terminating orbit for $\mu=0.9$, $\lambda=0.3$ (light gray region in Fig.~\ref{fig:6dDia}).]{
\includegraphics[width=0.31\textwidth]{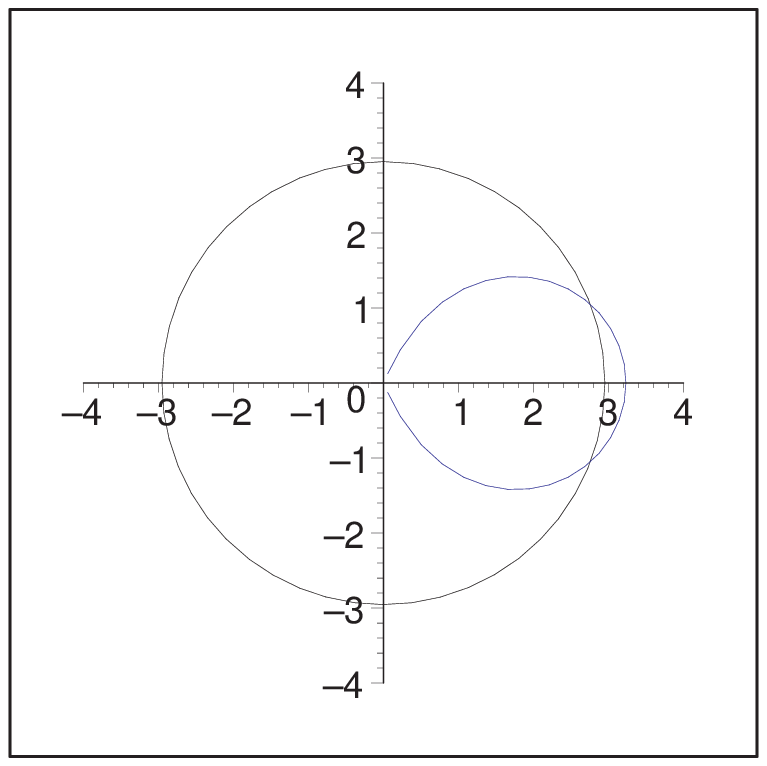}
}
\subfloat[][Infinite terminating orbit for $\mu=1.5$, $\lambda=0.6$ (white region in Fig.~\ref{fig:6dDia}).]{
\includegraphics[width=0.31\textwidth]{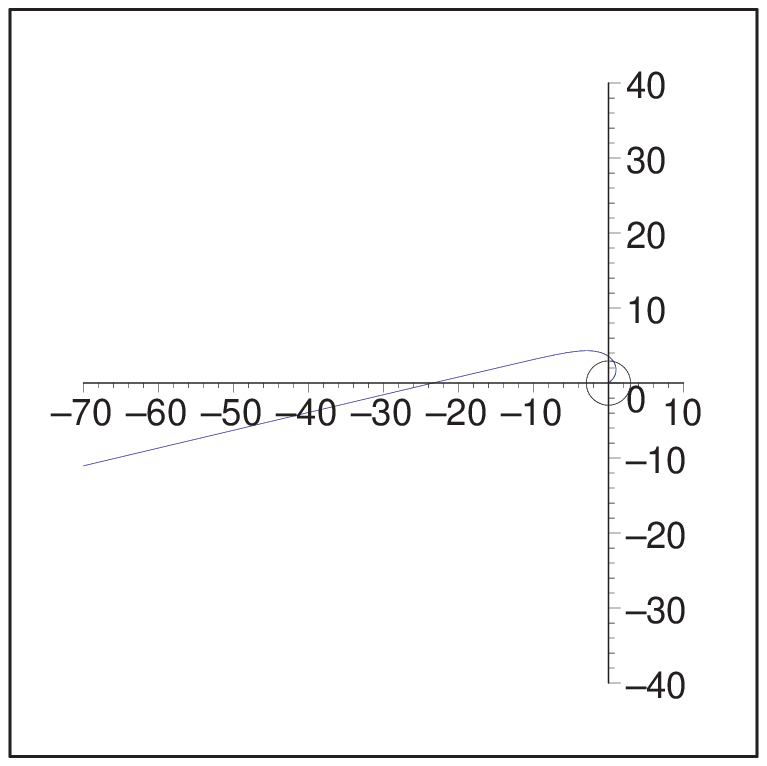}
}
\caption{Orbits for chosen values of $\mu$ and $\lambda$ in six--dimensional Schwarzschild space-time. The black circle indicates the Schwarzschild radius.}
\label{6dOrbits}
\end{figure*}

Figure \ref{fig:6dDia} shows the arrangement of zeros for the six--dimensional Schwarzschild space--time. The gray scale code is the same as in the Schwarzschild--(anti-)de Sitter case. There are no periodic bound orbits for any values of $\mu$ and $\lambda$ since the polynomial $P_6$ possesses at most two positive zeros. Some resulting orbits for chosen parameters $\mu$ and $\lambda$ from different regions in Fig.~\ref{fig:6dDia} are shown in Fig.~\ref{6dOrbits}. 

With our method it is also possible to analytically calculate the orbits of particles and light rays in Schwarzschild and Schwarzschild--(anti-)de Sitter space--times of up to 11 dimensions and in Reissner--Nordstr\"om--(anti-)de Sitter space--times of up to 7 dimensions. Corresponding work is in progress \cite{Hackmannetal08}. 

\section{Summary and outlook}

In this paper the explicit analytical solution for the geodesic motion of a point particle in a Schwarzschild--(anti-)de Sitter space--time has been presented. We were able to explicitly determine all possible solutions and to classify them. 

Analytic solutions are the starting point for approximation methods for the description of real stellar, planetary, comet, asteroid, or satellite trajectories
(see e.g. \cite{Hagihara70}). Analytic solutions of the geodesic equation can also serve as test beds for numerical codes for the dynamics of binary systems in the extreme stellar mass ratio case (extreme mass ratio inspirals, EMRIs) and also for the calculation of corresponding gravitational wave templates.

The methods presented here are not limited to the Schwarzschild--de Sitter case. They also can be applied to higher dimensional space--times like Schwarzschild-(anti-)de Sitter space--times of up to 11 dimensions and Reissner--Nordst\"om--(anti-)de Sitter space--times of up to 7 dimensions. Also for space--times as general as Pleba\'nski--Demia\'nski without acceleration the developed method can be applied \cite{Hackmannetal08a}. In this case polynomials of 6th order appear which will slightly complicate the structure of the orbits. It should be noted that this method, however, is not capable to solve equations of motions with an underlying polynomial of 7th or higher order. In such cases we have to enlarge the number of variables to be three or more. Then the Abel map is a mapping between 3 or higher dimensional spaces and it is not clear how to constrain this mapping in order to reduce the number of variables. 

\begin{acknowledgements}
We would like to thank H. Dittus, V. Kagramanova, J. Kunz, O. Lechtenfeld, D. Lorek and, in particular, P.H. Richter for fruitful discussions. Thanks are also due to W. Fischer for helping us with the calculations in Appendix B and due to the Pioneer Explorer Collaboration for assistance in Sec.~VI. Financial support of the German Aerospace Center DLR and of the German Research Foundation DFG is gratefully acknowledged. 
\end{acknowledgements}

\appendix

\section{Explicit computation of the analytical solution}

We describe now in detail the explicit computation of the analytical solution derived in Sec.~IV. According to \eqref{parameter} the parameters $\mu$, $\lambda$ and $\rho$ are related the energy $E$, angular momentum $L$ and cosmological constant $\Lambda$ which gives the polynomial $P_5$ in \eqref{Dgl_order5}. The zeros of $P_5$ have to be determined numerically using, e.g., a Newton method. The zeros of $P_5$ already characterize the type of orbit as described in Sec.~V. 

\begin{figure*}[t]
\subfloat[][5 real zeros $e_1$, ..., $e_5$]{
\includegraphics[width=0.28\textwidth]{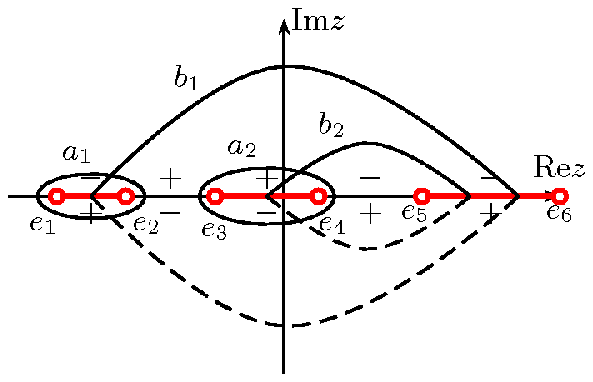}
}
\subfloat[][3 real zeros ($e_3$, $e_4$, $e_5$), 2 complex zeros ($e_1$, $e_2$)]{
\includegraphics[width=0.28\textwidth]{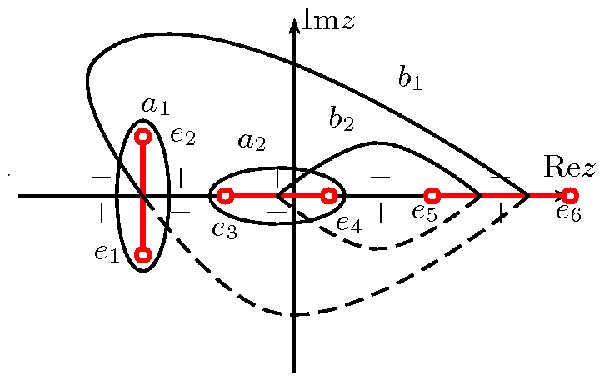}
}
\subfloat[][3 real zeros, 2 complex zeros]{
\includegraphics[width=0.28\textwidth]{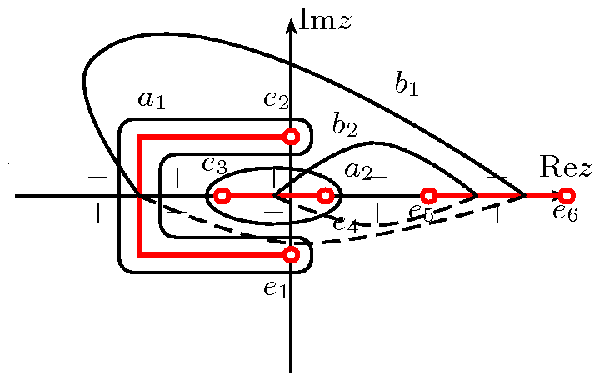}
} \\ 
\bigskip

\subfloat[][3 real zeros, 2 complex zeros]{
\includegraphics[width=0.28\textwidth]{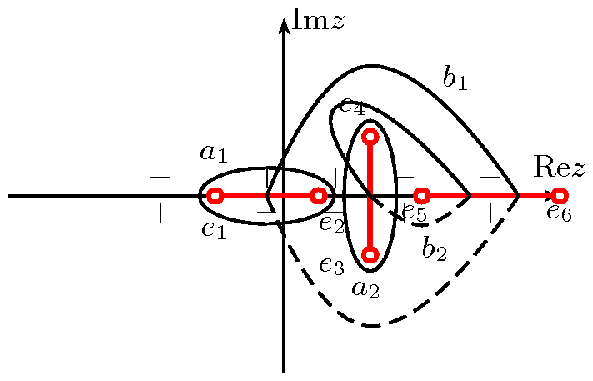}
} 
\subfloat[][3 real zeros, 2 complex zeros]{
\includegraphics[width=0.28\textwidth]{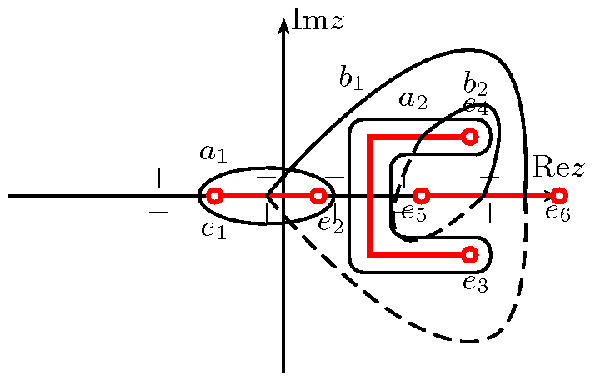}
}\\

\bigskip
\subfloat[][1 real zero ($e_5$), 4 complex zeros ($e_1$, ..., $e_4$)]{
\includegraphics[width=0.28\textwidth]{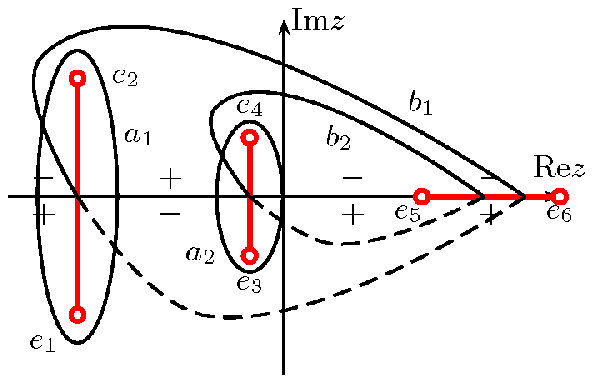}
} 
\subfloat[][1 real zero, 4 complex zeros]{
\includegraphics[width=0.28\textwidth]{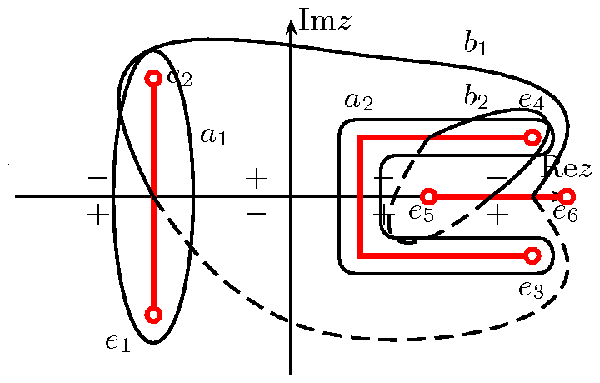}
}
\subfloat[][1 real zero, 4 complex zeros]{
\includegraphics[width=0.28\textwidth]{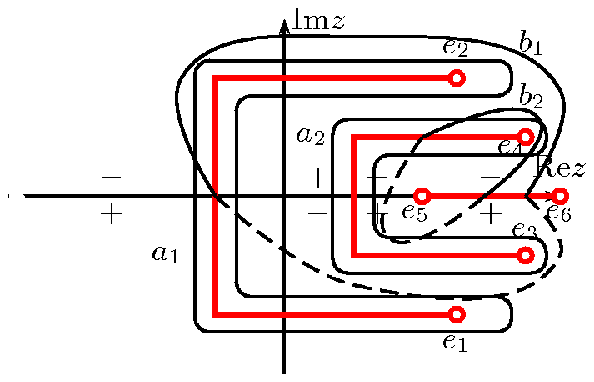}
}
\caption{Branch cuts (fat lines) and fundamental paths in the complex plane for all arrangements of zeros $e_1, \ldots, e_5$ of a polynomial of degree $5$ and $e_6 = \infty$. The branch cuts are drawn from $e_{2i-1}$ to $e_{2i}$. The completion of the $b$--paths on the other sheet is indicated by dashed lines.}
\label{fig:pathes}
\end{figure*}

The Riemannian surface corresponding to $\sqrt{P_5}$ is given by a two--torus, see Fig.~\ref{fig:brezel}. The fundamental paths $a_1, a_2$ and $b_1, b_2$ are always the same on this two-torus but change in the complex plane depending on the configuration of branch cuts, which have to chosen according to the zeros of $P_5$. There are eight  principally different arrangement of zeros of $P_5$ which are shown in Fig.~\ref{fig:pathes}. In each case we choose the branch cuts in such a way that we have a maximum number of pure real branch cuts. The branch cuts and the resulting canonically fundamental paths are also shown in Fig.~\ref{fig:pathes} together with the appropriate sign of the square root on one sheet (with reversed sign on the other sheet).

Next we have to calculate the period matrices $(2\omega, 2\omega')$ and $(2\eta, 2\eta')$ given in \eqref{periodmatrices} and, from that, the normalized period matrix $\tau$ defined in \eqref{normalizedtau}. The periods corresponding to paths $a_i$, $i=1,2$, can be obtained by integration from $e_{2i-1}$ to $e_{2i}$; the integration back from $e_{2i}$ to $e_{2i-1}$ yields just the same value due to the different sign of the square root. The periods corresponding to paths $b_j$, $j=1,2$, can be calculated by integrating from $e_{2j}$ to $e_{2j+1}$. For paths which encircle one or two complex zeros, the most convenient way to calculate the integral is to take a path running straight from the complex zero to the real axis and then to proceed along the real axis. Of course, fundamental paths can be distorted but only in such a way that they do not cross other branch cuts or paths as the one showed in Fig.~\ref{fig:pathes}. 

The branch points $e_i$ are always singularities of the integrand and, thus, may cause numerical problems. These problems can be handled by a partial integration of the type 
\begin{equation}
\int_{e_i}^{p} \frac{dx}{\sqrt{\prod_{j=1}^5 (x-e_j)}} = 2 \left. \frac{\sqrt{x-e_i}}{\sqrt{Q(x)}} \right|_{e_i}^p - \int_{e_i}^p \frac{\sqrt{x-e_i}}{\sqrt{Q(x)}^3} Q'(x) dx \,, 
\end{equation}
where $Q(x) = \prod_{j \neq i} (x-e_j)$ and $p$ is some point on the integration path. As a consequence, the resulting integrand is no longer singular. In the case that all five roots of $P_5$ are real, the computation then is straightforward since $P_5$ is always real on the integration paths. If some roots are complex, the integration is more involved and one has to take into account that the real part of $P_5$ is symmetric with respect to the real axis while the imaginary part is antisymmetric. One also has to carefully choose the appropriate branch of the square root.

After having calculated all the periods we choose initial values $u_0$ and $\varphi_0$ and determine $\varphi'_0 = \varphi_0 + \int_{u_0}^\infty dz_2$. Note that the integral is just a sum of the half--periods $(\omega,\omega')$ already calculated above provided $u_0$, as usual, is taken to be one of the zeros of $P_5$. For every $\varphi$ we have to find now the dummy parameter $\varphi_1$ such that $(2 \omega)^{-1} \vec \varphi_\Theta$ is an element of the theta divisor $\Theta_{\vec K_\infty} = \left\{ z \,|\, \vartheta \left[ \vek{1/2}{1/2},\vek{0}{1/2} \right] (z;\tau) = 0\right\}$. This dummy parameter has no physical meaning and depends, beside $\varphi$, on the initial values $u_0$, $\varphi_0$ and the normalized period matrix $\tau$. The value of $\varphi_1$ can be computed with a Newton method for the function $g \circ h: \mathbbm{C} \to \mathbbm{C}$, $\varphi \overset{h}{\mapsto} (2\omega)^{-1} \vek{\varphi_1}{\varphi-\varphi'_0} \overset{g}{\mapsto} \vartheta \left[ \vek{1/2}{1/2}, \vek{0}{1/2} \right] (z;\tau)$. For $\varphi = \varphi_0$ and $u_0$ being a branch point, the dummy parameter $\varphi_1$ is explicitly known. From $\varphi - \varphi'_0 = \int_\infty^{u_0} dz_2$ and $\vartheta \left[ \vek{1/2}{1/2}, \vek{0}{1/2} \right] (z;\tau) = 0$ for $z=\tau n+m$, $n,m \in \mathbbm{Z}$, we obtain in this case $\varphi_1 = \int_\infty^{u_0} dz_1$.    

After these calculations we are finally able to compute 
\begin{equation}
r(\varphi) = \dfrac{r_{\rm S}}{u(\varphi)} = - r_{\rm S} \dfrac{\sigma_2(\vec \varphi_\Theta)}{\sigma_1(\vec \varphi_\Theta)} \,.
\end{equation} 
The solution $r(\varphi)$ can be computed pointwise for any value of $\varphi$ with, in principle, arbitrary accuracy. 

\section{Calculation of post--Schwarzschild period}

Toward an integration of the second term of \eqref{taylor} we first obtain
\begin{widetext}
\begin{align}
\oint_{a_i} \frac{x^2+\lambda}{x^2 P_3(x) \sqrt{P_3(x)}} dx & = \oint_A \frac{(4z+\frac{1}{3})^2 + \lambda}{(4z+\frac{1}{3})^2 \cdot 4^2 P_W(z) \sqrt{4^2 P_W(z)}} 4dz \nonumber \\
& = \frac{1}{4^2} \left( \oint_A \frac{dz}{P_W(z) \sqrt{P_W(z)}} + \lambda \oint_A \frac{dz}{(4z+\frac{1}{3})^2 P_W(z) \sqrt{P_W(z)}} \right)\,. \label{Lambdapart}
\end{align}
\end{widetext}
In the following we will represent the functions
\begin{equation}
F_1(z) = \frac{1}{P_W(z)}, \quad F_2(z) = \frac{1}{(4z+\frac{1}{3})^2 P_W(z)}\,.
\end{equation}
as linear combinations of the Weierstrass elliptic function $\wp$ as well as the Weierstrass $\zeta$ function. The reason is that these functions can be integrated easily since $\zeta' = -\wp$ and $(\log \sigma)' = \zeta$, where $\sigma$ is the Weierstrass' $\sigma$--function:
\begin{align}
\int_\gamma \wp(u-u_0) du & = \zeta(\gamma(0) - u_0) - \zeta(\gamma(1) - u_0) \label{int_zeta}\\
\int_\gamma \zeta(u-u_0) du & = \log \sigma(\gamma(1) - u_0) - \log \sigma(\gamma(0) - u_0) \,, \label{int_sigma}
\end{align} 
where $\gamma(0) = u_1$ and $\gamma(1) = u_1+\omega_1$ as above (the branches of $\log$ will be discussed later). For the $\zeta$-- and $\sigma$--functions we have a quasiperiodicity
\begin{align}
\zeta(u+\omega_j) & = \zeta(u) + \eta_j \\
\sigma(u+\omega_j) & = e^{\eta_j(u+\omega_j/2) + \pi i} \sigma(u) \, ,
\end{align}
where $\eta_j$ are periods of second kind given by 
\begin{equation}
\begin{split}
\eta_1 & = - \oint_A \frac{z dz}{\sqrt{P_W(z)}} = - 2 \int_{\rho_3}^{\rho_2}  \wp(u) du \\ 
\eta_2 & = - \oint_B \frac{z dz}{\sqrt{P_W(z)}} = -2 \int_{\rho_2}^{\rho_1} \wp(u) du \, . 
\end{split}
\end{equation}
Thus, \eqref{int_zeta} and \eqref{int_sigma} can be rewritten as
\begin{align}
\int_\gamma \wp(u-u_0) du & =  -  \eta_1\\
\int_\gamma \zeta(u-u_0) du & =  \eta_1(u_1-u_0+ \tfrac{1}{2} \omega_1) + \pi i + 2\pi i k
\end{align}
with $k \in \mathbbm{Z}$. 

\subsection{Integration of $F_1/\sqrt{P_W}$:}

First we substitute $z=\wp(u)$
\begin{equation}
F_1(z) = \frac{1}{P_W(z)} = \frac{1}{P_W(\wp(u))} = \frac{1}{\wp'(u)^2}=: f_1(u)\,.
\end{equation}
The function $f_1$ only possesses poles of second order in $\rho_1$, $\rho_2$, and $\rho_3$. In a neighborhood of $\rho_j$, the function $f_1$ can be expanded as
\begin{equation}\label{f1}
f_1(u) = \frac{a_{j2}}{(u-\rho_j)^2} + \frac{a_{j1}}{u-\rho_j} + \text{holomorphic part} \,.
\end{equation}
Since $\wp'(\rho_j+z)^2=\wp'(\rho_j-z)^2$ for all $j$ and $z$, $f_1$ is symmetric with respect to all $\rho_j$ and, therefore, depends only on even powers of $(u - \rho_j)$ so that $a_{j1} = 0$. The constant $a_{j2}$ can be evaluated with a comparison of coefficients. For this, we note that $\wp'(\rho_j) = 0 = \wp'''(\rho_j)$ and, thus,
\begin{equation}\label{wp'}
\wp'(u) = \wp''(\rho_j) (u-\rho_j) + \sum_{i=3}^\infty c_i (u-\rho_j)^i
\end{equation}
in a neighborhood of $\rho_j$ and for some constants $c_i$. If we square both sides of the equation, we see that $\wp'^2$ contains only even powers of $(u-\rho_j)$ larger than 1. It follows
\begin{equation}
1 = f_1(u) \wp'(u)^2 = a_{j2} \wp''(\rho_j)^2 + \text{higher powers of } (u-\rho_j) \,.
\end{equation}
From that it follows $a_{j2} = \frac{1}{\wp''(\rho_j)^2}$ for all $j$. The function $\wp(u-\rho_j)$ has only one pole of second order in $\rho_j$ with zero residue. Therefore, the difference 
\begin{equation}
f_1(u) - \sum_{j=1}^3 a_{j2} \wp(u-\rho_j)
\end{equation}
is a holomorphic elliptic function and, thus, is constant. This yields
\begin{equation}
f_1(u) = \sum_{j=1}^3 a_{j2} \wp(u-\rho_j) + c_1\,.
\end{equation}
The constant $c_1$ can be determined by $f_1(0) = 0$ using the relation $\wp(- \rho_j) = \wp(\rho_j) = z_j$:
\begin{equation}
c_1 = - \sum_{j=1}^3 a_{j2} z_j\,.
\end{equation}

In summary, we obtain
\begin{align}
\oint_A \frac{dz}{P_W(z) \sqrt{P_W(z)}} & = \int_\gamma f_1(u) du \nonumber\\
& = \int_\gamma \sum_{j=1}^3 a_{j2} ( \wp(u-\rho_j) - z_j) du \nonumber \\
& = \sum_{j=1}^3 \frac{1}{\wp''(\rho_j)^2} \left( \int_\gamma \wp(u-\rho_j) du - z_j \omega_1 \right) \nonumber \\
& = \sum_{j=1}^3 \frac{1}{\wp''(\rho_j)^2} \left( -\eta_1 - z_j \omega_1\right) \,. \label{F1}
\end{align}

\subsection{Integration of $F_2/\sqrt{P_W}$:}

Again, we first substitute $z=\wp(u)$ and obtain
\begin{align}
F_2(z) & = \frac{1}{(4z+\frac{1}{3})^2 P_W(z)} = \frac{1}{(4\wp(u)+\frac{1}{3})^2 P_W(\wp(u))}\nonumber\\
& = \frac{1}{((4\wp(u)+\frac{1}{3}) \wp'(u) )^2}=: f_2(u)\,.
\end{align}
The function $f_2$ possesses poles of second order in $\rho_1$, $\rho_2$, $\rho_3$ and in all $u_0 \in R$ such that $\wp(u_0) = - \frac{1}{12}$. Since we assumed that the considered orbit is bound, all zeros of $P_3$ have to be positive and, thus, $z_1>z_2>z_3>- \frac{1}{12}$. This means that $0<\Im(u_0)<\Im(\rho_3)$. The function $\wp$ is even and, hence, also $\tilde{u}_0 := \omega_2 - u_0 \in R$ is a pole of second order (see Fig.~\ref{fig:rectangle}).

Since $\wp$ is symmetric with respect to $\rho_j$, the function $f_2$ can be expanded in the same way as above as
\begin{equation}
f_2(u) = \frac{a_{j2}}{(u-\rho_j)^{2}} + \text{holomorphic part}
\end{equation}
in a neighborhood of $\rho_j$. An expansion of $(4\wp(u)+\tfrac{1}{3}) \wp'(u)$ near $\rho_j$ yields
\begin{align}
\left(4\wp(u)+\tfrac{1}{3}\right) \wp'(u) & = \alpha_{j1} (u-\rho_j) + \alpha_{j2} (u-\rho_j)^2 \nonumber\\
& \quad + \text{higher order terms}
\end{align}
because of $\wp'(\rho_j)=0$. The coefficients are given by
\begin{align}
\alpha_{j1} & = \left( \left( 4\wp(u)+\tfrac{1}{3} \right) \wp'(u) \right)'_{u=\rho_j} \nonumber\\
& = \left(4z_j + \tfrac{1}{3}\right) \wp''(\rho_j)\\
\alpha_{j2} & = \left( \left( 4\wp(u)+\tfrac{1}{3} \right) \wp'(u) \right)''_{u=\rho_j} = 0 \, .
\end{align}
A comparison of coefficients
\begin{align}
1 & = f_2(u) \left( \left( 4\wp(u)+\tfrac{1}{3} \right) \wp'(u) \right)^2 \nonumber\\
& = a_{j2} \alpha_{j1}^2 + \text{higher powers of } (u-\rho_j)
\end{align}
yields
\begin{align}
f_2(u) & = \frac{1}{(u-\rho_j)^{2}} \left( \left( 4z_j+\tfrac{1}{3} \right) \wp''(\rho_j) \right)^{-2} \nonumber\\
& \quad + \text{holomorphic part} \,.
\end{align}
in a neighborhood of $\rho_j$.

The same procedure will be carried through for $u_0$ and $\tilde{u}_0$. We have
\begin{equation}
f_2(u) = \frac{b_{2}}{(u-u_0)^{2}} + \frac{b_{1}}{u-u_0} + \text{holomorphic part}
\end{equation}
and
\begin{align}
\left(4\wp(u)+\tfrac{1}{3}\right) \wp'(u) & = \beta_{1} (u-u_0) + \beta_{2} (u-u_0)^2 \nonumber\\
& \quad + \text{higher order terms} \label{u_0}
\end{align}
near $u_0$. The coefficients of \eqref{u_0} read
\begin{align}
\beta_{1} & = 4\wp'(u_0)^2 \\
\beta_2 & = 6 \wp'(u_0) \wp''(u_0)\,.
\end{align} 
Again, a comparison of coefficients
\begin{align}
1 & = f_2(u) \left( \left(4\wp(u)+\tfrac{1}{3}\right) \wp'(u) \right)^2 \nonumber\\
& =   b_2 \beta_1^2 + (2 b_2 \beta_1\beta_2+b_1\beta_1^2) (u-u_0) \nonumber\\
& \quad + \text{higher order terms}
\end{align}
yields
\begin{align}
b_2 & = \beta_1^{-2} = \frac{1}{16 \wp'(u_0)^{4}}\\
b_1 & = -2\beta_1\beta_2b_2 \beta_1^{-2} = -2\beta_2 \beta_1^{-3} = - \frac{3}{16} \frac{\wp''(u_0)}{\wp'(u_0)^5} \,.
\end{align}

In a neighborhood of $\tilde{u}_0$, the function $f_2$ is given by
\begin{equation}
f_2(u) = \frac{\tilde{b}_{2}}{(u-\tilde{u}_0)^{2}} + \frac{\tilde{b}_{1}}{u-\tilde{u}_0} + \text{holomorphic part} \,.
\end{equation}
As $\wp'$ is an odd and $\wp''$ an even function we get for the coefficients of the expansion of $(4\wp(u)+\frac{1}{3}) \wp'(u)$ near $\tilde{u_0}$ with $\tilde{u_0} = \omega_2-u_0$ the relations
\begin{equation}
\tilde{\beta}_1 = \beta_1, \quad \tilde{\beta}_2 = - \beta_2
\end{equation}
and, therefore, 
\begin{equation}
\tilde{b}_1 = - b_1, \quad \tilde{b}_2 = b_2\,. 
\end{equation}

Summarized, the function
\begin{align}
g_2(u) := & \sum_{j=1}^3 a_{j2} \wp(u-\rho_j) + b_2 ( \wp(u-u_0) + \wp(u-\tilde{u}_0) ) \nonumber\\
&  + b_1 ( \zeta(u-u_0) - \zeta(u-\tilde{u}_0) )
\end{align}
has the same poles with the same coefficients as $f_2$. Therefore, $f_2 - g_2$ is a holomorphic elliptic function and, thus, is equal to a constant $c_2$. This constant can be determined by the condition $0 = f_2(0) = g_2(0)+c_2$ which yields
\begin{equation}
c_2 = - \sum_{j=1}^3 a_{j2} z_j +\tfrac{1}{6} b_2 - b_1 (\zeta(\tilde{u}_0) - \zeta(u_0) )\,.
\end{equation}
As a consequence,
\begin{align}
f_2(u) & = \sum_{j=1}^3 a_{j2} (\wp(u-\rho_j) - z_j) \nonumber\\
& \quad + b_2 ( \wp(u-u_0) + \wp(u-\tilde{u}_0) + \tfrac{1}{6} ) \nonumber\\
& \quad + b_1 ( \zeta(u-u_0) - \zeta(u-\tilde{u}_0) + \zeta(u_0) - \zeta(\tilde{u}_0) )\,.
\end{align}

Now we can carry through the integration of the second term on the right-hand side of \eqref{Lambdapart}:
\begin{widetext}
\begin{align}
\oint_A F_2(z) \frac{dz}{\sqrt{P_W(z)}} & = \int_\gamma f_2(u) du \nonumber \\
& = \sum_{j=1}^3 a_{j2} \left( \int_\gamma \wp(u-\rho_j) du - z_j \omega_1\right) + b_2 \int_\gamma \wp(u-u_0) + \wp(u-\tilde{u}_0) du + \tfrac{1}{6} \omega_1 \nonumber  \\
& \quad + b_1 \int_\gamma \zeta(u-u_0) - \zeta(u-\tilde{u}_0) du + b_1 \omega_1 (\zeta(u_0) - \zeta(\tilde{u}_0) ) \nonumber \\
& = - \sum_{j=1}^3 a_{j2} ( \eta_1 + z_j \omega_1) + b_2 \left( \tfrac{1}{6} \omega_1 - 2\eta_1 \right) + b_1 ( \eta_1 ( \tilde{u}_0 - u_0 ) + 2\pi i (k_1-k_2) ) \nonumber \\
& \quad + b_1 \omega_1 ( \zeta(u_0) - \zeta(\tilde{u}_0) ) \nonumber  \\
& = - \sum_{j=1}^3  a_{j2} ( \eta_1 + z_j \omega_1) + b_2 \left( \tfrac{1}{6} \omega_1 - 2\eta_1\right)  \nonumber \\ 
& \quad + b_1 ( \eta_1 (\omega_2 - 2u_0) + 2\pi i (k_1-k_2) ) + b_1 \omega_1 ( 2\zeta(u_0) - \eta_2) \,.
\end{align}
\end{widetext}

The difference $(k_1-k_2)$ can be calculated as follows. First note that via $x = 4\wp(u)+\frac{1}{3}$, $u_0$ corresponds to $0$ and $u_1$ to $x_3$. Since $0<x_3$ for  bound orbits under consideration we have $\Im(u_0) < \Im(u_1) < \Im(\tilde{u}_0)$. Let now $l$ be determined by
\begin{align}
\int_{u_2}^{u_2+\omega_1} \zeta(u-\tilde{u}_0) du & = \eta_1 (u_2 - \tilde{u}_0+\frac{\omega_1}{2} ) + \pi i + 2\pi i l \,,
\end{align}
where $u_2 \in i \cdot \mathbbm{R}$ is such that $\Im(u_2) > \Im(\tilde{u}_0) > \Im(u_1) > \Im(u_0)$. From
\begin{itemize}
\item[{(i)}] $l$ does not depend on $\tilde{u}_0$ as long as $\Im(\tilde{u}_0) < \Im(u_2)$ holds and, thus, 
\begin{equation}\label{u_2}
\int_{u_2}^{u_2+\omega_1} \zeta(u-u_0) du = \eta_1 (u_2 - u_0+\frac{\omega_1}{2} ) + \pi i + 2\pi i l
\end{equation}
and
\item[{(ii)}] \eqref{u_2} holds also for $u_2$ replaced by $u_1$ by Cauchy's integral formula for the rectangle with corners $u_1$, $u_1+\omega_1$, $u_2+\omega_1$ and $u_2$.
\end{itemize}
it follows that $l=k_1$.

We show now that $k_2=l+1$ and, thus, $k_2=k_1+1$. Consider the counterclockwise oriented rectangle with corners $u_1$, $u_1+\omega_1$, $u_2+\omega_1$ and $u_2$. Let $c$ be the boundary of this rectangle but with a two symmetric small bumps such that $c$ encircles the pole $\tilde{u}_0$ of $\zeta(u-\tilde{u}_0)$ with residue $1$, but not $\tilde{u}_0+\omega_1$. Then the residue theorem gives
\begin{widetext}
\begin{align}
2\pi i & = \oint_c \zeta(u-\tilde{u}_0) du \nonumber  \\
& = \int_{u_1}^{u_1+\omega_1} \zeta(u-\tilde{u}_0) du + \int_{u_1+\omega_1}^{u_2+\omega_1} \zeta(u-\tilde{u}_0) du + \int_{u_2+\omega_1}^{u_2} \zeta(u-\tilde{u}_0) du + \int_{u_2}^{u_1} \zeta(u-\tilde{u}_0) du \nonumber \\
& = \int_{u_1}^{u_1+\omega_1} \zeta(u-\tilde{u}_0) du - \int_{u_2}^{u_2+\omega_1} \zeta(u-\tilde{u}_0) du + \eta_1(u_2-u_1) \nonumber \\
& = \eta_1 (u_1-\tilde{u}_0+ \tfrac{1}{2} \omega_1) + \pi i + 2\pi i k_2 -  ( \eta_1 (u_2 - \tilde{u}_0+ \tfrac{1}{2} \omega_1) + \pi i + 2\pi i l ) + \eta_1(u_2-u_1) \nonumber \\
& = 2\pi i (k_2-l)\,.
\end{align}

With the Legendre relation $\eta_1 \omega_2 - \eta_2 \omega_1 = 2\pi i$ we finally obtain
\begin{equation}\label{F2}
\oint_A F_2(z) \frac{dz}{\sqrt{P_W(z)}} =  - \sum_{j=1}^3 \left( \frac{\eta_1 + z_j \omega_1}{\left( 4z_j+\frac{1}{3} \right)^2 \wp''(\rho_j)^2} \right) + \frac{\frac{1}{6} \omega_1 - 2\eta_1}{16 \wp'(u_0)^{4}} - \frac{6}{16} \frac{\wp''(u_0)}{\wp'(u_0)^5} (- \eta_1 u_0 + \omega_1 \zeta(u_0))\,.
\end{equation}
Note that though the values $\wp'(u_0)^5$, $u_0$ and $\zeta(u_0)$ appearing in the last part of the right-hand side are all purely imaginary, the hole term is real.
\end{widetext}

\bibliographystyle{unsrt}
\bibliography{SdS}

\begin{thebibliography}{10}

\bibitem{Will93}
C.M. Will.
\newblock {\em Theory and Experiment in Gravitational Physics (Revised
  Edition)}.
\newblock Cambridge University Press, Cambridge, 1993.

\bibitem{Will01}
C.M. Will.
\newblock The confrontation between general relativity and experiment.
\newblock {\em Living Rev. Relativity}, 2001, ww.livingreviews.org/lrr-2001-4.

\bibitem{Krameretal06a}
M.~Kramer, I.H. Stairs, R.N. Manchester, M.A. MacLaughlin, A.G. Lyre, R.D.
  Ferdman, M.~Burgag, D.R. Lorimer, A~Possenti, N.~D'Amico, J.~Sarkission, G.B.
  Hobbs, J.E. Reynolds, P.C.C. Freire, and F.~Camilo.
\newblock Tests of general relativity from timing the double pulsar.
\newblock {\em Science}, 314:97, 2006.

\bibitem{KagramanovaKunzLaemmerzahl06}
V.~Kagramanova, J.~Kunz, and C.~L{\"a}mmerzahl.
\newblock Solar system effects in {S}chwarzschild--de {S}itter space--time.
\newblock {\em Phys. Lett.}, A 634:465, 2006.

\bibitem{JetzerSereno06}
P.~Jetzer and M.~Sereno.
\newblock Two-body problem with the cosmological constant and observational
  constraints.
\newblock {\em Phys. Rev.}, D 73:044015, 2006.

\bibitem{KerrHauckMashhoon03}
A.W. Kerr, J.C. Hauck, and B.~Mashhoon.
\newblock Standard clocks, orbital precession and the cosmological constant.
\newblock {\em Class. Qauntum Grav.}, 20:2727, 2003.

\bibitem{Andersonetal02}
J.D. Anderson, P.A. Laing, E.L. Lau, A.S. Liu, M.M. Nieto, and S.G. Turyshev.
\newblock Study of the anomalous acceleration of {P}ioneer 10 and 11.
\newblock {\em Phys. Rev.}, D 65:082004, 2002.

\bibitem{BalagueraBoehmerNoeakowski06}
A.~Balaguera-Antolinez, C.G. B{\"o}hmer, and M.~Nowakowski.
\newblock Scales set by the cosmological constant.
\newblock {\em Class. Quantum Grav.}, 23:485, 2006.

\bibitem{KunduriLucietti05}
H.K. Kunduri and J.~Lucietti.
\newblock Integrability and the {K}err-({A})d{S} black hole in five dimensions.
\newblock {\em Phys. Rev.}, D 71:104021, 2005.

\bibitem{VasudevanStevensPage05}
M.~Vasudevan, K.A. Stevens, and D.N. Page.
\newblock Separability of the {H}amilton{J}acobi and {K}lein{G}ordon equations
  in {K}errde {S}itter metrics.
\newblock {\em Class. Quantum Grav.}, 22:339, 2005.

\bibitem{Vasudevan05}
M.~Vasudevan.
\newblock Integrability of some charged rotating supergravity black hole
  solutions in four and five dimensions.
\newblock {\em Phys. Lett.}, B 624:287, 2005.

\bibitem{Chongetal05}
Z.-W. Chong, M.~Cvetic, H.~Lu, and C.N. Pope.
\newblock General non-extremal rotating black holes in minimal five-dimensional
  gauged supergravity.
\newblock {\em Phys. Rev. Lett.}, 95:161301, 2005.

\bibitem{Pageetal07}
D.N. Page, D.~Kubiz{\v{n}}{\'a}k, M.~Vasudevan, and P.~Krtou{\v{s}}.
\newblock Complete integrability of geodesic motion in general
  higher-dimensional rotating black-hole spacetimes.
\newblock {\em Phys. Rev. Lett.}, 98:061102, 2007.

\bibitem{Hagihara31}
Y.~Hagihara.
\newblock Theory of relativistic trajectories in a gravitational field of
  {S}chwarzschild.
\newblock {\em Japan. J. Astron. Geophys.}, 8:67, 1931.

\bibitem{Chandrasekhar83}
S.~Chandrasekhar.
\newblock {\em The Mathematical Theory of Black Holes}.
\newblock Oxford University Press, Oxford, 1983.

\bibitem{HackmannLaemmerzahl08}
E.~Hackmann and C.~L{\"a}mmerzahl.
\newblock Complete analytic solution of the geodesic equation in
  schwarzschild--(anti) de sitter space--times.
\newblock {\em Phys. Rev. Lett.}, 100:171101--1, 2008.

\bibitem{CruzOlivaresVillanueva05}
N.~Cruz, M.~Olivares, and J.R. Villanueva.
\newblock The geodesic structure of the {S}chwarzschild anti-de {S}itter black
  hole.
\newblock {\em Class. Quantum Grav.}, 22:1167, 2005.

\bibitem{Abel1828}
N.H. Abel.
\newblock Remarques sur quelques properietes generales d'une certaine sorte de
  fonctions transcendentes.
\newblock {\em Crelle's J. Math.}, 3:313, 1828.

\bibitem{Jacobi1832}
C.~G.~J. Jacobi.
\newblock Considerationes generales de transcendentibus abelianis.
\newblock {\em Crelle's J. Math.}, 9:394, 1832.

\bibitem{Weierstrass1854}
K.~Weierstrass.
\newblock Zur theorie der abelschen functionen.
\newblock {\em Crelle's J. Math.}, 47:289, 1854.

\bibitem{Baker1895}
H.F. Baker.
\newblock {\em Abelian Functions. Abel's theorem and the allied theory of theta
  functions}.
\newblock Cambridge University Press, Cambridge, 1995. First published 1897.

\bibitem{KraniotisWhitehouse03}
G.V. Kraniotis and S.B. Whitehouse.
\newblock Compact calculation of the perihelion precession of mercury in
  general relativity, the cosmological constant and {J}acobis inversion
  problem.
\newblock {\em Class. Quantum Grav.}, 20:4817, 2003.

\bibitem{Drociuk02}
R.~Drociuk.
\newblock Cosmic force, 2002.
\newblock gr-qc/0204023.

\bibitem{Kraniotis04}
G.V. Kraniotis.
\newblock Precise relativistic orbits in {K}err and {K}err--(anti) de {S}itter
  spacetimes.
\newblock {\em Class. Quantum Grav.}, 21:4743, 2004.

\bibitem{EnolskiiPronineRichter03}
V.Z. Enolskii, M.~Pronine, and P.H. Richter.
\newblock Double pendulum and $\theta$-divisor.
\newblock {\em J. Nonlinear Sc.}, 13:157, 2003.

\bibitem{Rindler01}
W.~Rindler.
\newblock {\em Relativity}.
\newblock Oxford University Press, Oxford, 2001.

\bibitem{Geyer80}
K.H. Geyer.
\newblock Geometrie der {R}aum--{Z}eit der {M}a{\ss}bestimmung von {K}ottler,
  {W}eyl und {T}refftz.
\newblock {\em Astr. Nachr.}, 301:135, 1980.

\bibitem{Miranda95}
R.~Miranda.
\newblock {\em Algebraic Curves and Riemann Surfaces}.
\newblock American Math. Soc., Providence, 1995.

\bibitem{RauchFarkas74}
H.E. Rauch and H.M. Farkas.
\newblock {\em Theta Functions with Applications to Riemann Surfaces}.
\newblock Williams and Wilkins, Baltimore, 1974.

\bibitem{BuchstaberEnolskiiLeykin97}
V.M. Buchstaber, V.Z. Enolskii, and D.V. Leykin.
\newblock {\em Hyperelliptic Kleinian Functions and Applications}.
\newblock Reviews in Mathematics and Mathematical Physics 10. Gordon and
  Breach, 1997.

\bibitem{Mumford83}
D.~Mumford.
\newblock {\em Tata Lectures on Theta, Vol. I and II}.
\newblock Birkh{\"a}user, Boston, 1983/84.

\bibitem{Note1}
The symbol $\protect \frac {1}{2} \protect \mathbbm {Z}^g$ denotes the set of
  all $g$-dimensional vectors with half--integer entries $\protect \ldots ,
  -\protect \frac {3}{2}, -1, -\protect \frac {1}{2}, 0, \protect \frac {1}{2},
  1, \protect \frac {3}{2}, \protect \ldots $.

\bibitem{GradshteynRyzhik83}
I.S. Gradshteyn and I.M. Ryzhik.
\newblock {\em Table of Integrals, Series, and Products}.
\newblock Academic Press, Orlando, 1983.

\bibitem{AbramowitzStegun68}
M.~Abramowitz and I.A.~(Ed.) Stegun.
\newblock {\em Handbook of mathematical functions}.
\newblock Dover Publications, New York, 1968.

\bibitem{RindlerIshak07}
W.~Rindler and M.~Ishak.
\newblock Contribution of the cosmological constant to the relativistic bending
  of light revisited.
\newblock {\em Phys. Rev.}, D 76:043006, 2007.

\bibitem{NietoAnderson05}
M.M. Nieto and J.D. Anderson.
\newblock Using early data to illuminate the {P}ioneer anomaly.
\newblock {\em Class. Quantum. Grav.}, 22:5343, 2005.

\bibitem{Hurwitz64}
A.~Hurwitz.
\newblock {\em Vorlesungen \"uber Allgemeine Funktionentheorie und elliptische
  Funktionen}.
\newblock Springer--Verlag, Berlin, 1964.

\bibitem{Note2}
http://history.nasa.gov/SP-423/intro.htm.

\bibitem{Valtonen08}
M.J. Valtonen.
\newblock {OJ}287: A binary black hole system.
\newblock {\em RevMexAA}, 32:22, 2008.

\bibitem{Valtonenetal08}
M.J. Valtonen, H.J. Lehto, K~Nilsson, J.~Heidt, L.O. Takalo, A.~Sillanpaa,
  C.~Villforth, M.~Kidger, G.~Poyner, T.~Pursimo, S.~Zola, J.~H. Wu, X.~Zhou,
  K.~Sadakane, M.~Drozdz, D.~Koziel, D.~Marchev, W.~Ogloza, C.~Porowski,
  M.~Siwak, G.~Stachowski, M.~Winiarski, V.P. Hentunen, M.~Nissinen, A.~Liakos,
  and S.~Dogru.
\newblock A massive binary black hole system in {OJ}287 and a test of general
  relativity.
\newblock {\em Nature}, 452:851, 2008.

\bibitem{Tangherlini63}
F.R. Tangherlini.
\newblock Schwarzschild field in {$n$} dimensions and the dimensionality of
  space problem.
\newblock {\em Nuovo Cim.}, 27:636, 1963.

\bibitem{Hackmannetal08}
E.~Hackmann, V.~Kagramanova, J.~Kunz, and C.~L{\"a}mmerzahl.
\newblock Analytic solutions of the geodesic equations in higher dimensional
  space--times.
\newblock {\em in preparation}, 2008.

\bibitem{Hagihara70}
Y.~Hagihara.
\newblock {\em Celestial Mechanics}.
\newblock MIT Press, Cambridge, Mass., 1970.

\bibitem{Hackmannetal08a}
E.~Hackmann, V.~Kagramanova, J.~Kunz, and C.~L{\"a}mmerzahl.
\newblock Analytic solution of the geodesic equation in
  {P}lebanski--{D}emianski space-time in terms of hyperelliptic functions.
\newblock {\em in preparation}, 2008.

\end{thebibliography}

\end{document}